\documentclass[]{aa}
\usepackage[T1]{fontenc}
\usepackage[utf8]{inputenx}
\usepackage{lmodern}
\usepackage{graphicx}
\usepackage{natbib}
\bibpunct{(}{)}{;}{a}{}{,} % to follow the A&A style
\setcitestyle{round}
\usepackage{ulem}
\usepackage{textcomp}
\usepackage{gensymb}
\usepackage{longtable}
\usepackage{threeparttable}
\usepackage{multicol}
\usepackage{multirow}
\usepackage{lipsum}
\usepackage{float}
\usepackage{todonotes}
\usepackage[Symbol]{upgreek}
\usepackage{amsmath}
\usepackage{etoolbox}
\usepackage{txfonts}
\usepackage{url}
\usepackage{xcolor}
\usepackage{mathtools}
\usepackage[most]{tcolorbox}
\usepackage{xcolor}
\usepackage[colorlinks=true,allcolors=blue]{hyperref}
\usepackage{adjustbox}

\usepackage{array}
\newcolumntype{L}[1]{>{\raggedright\let\newline\\\arraybackslash\hspace{0pt}}m{#1}}
\newcolumntype{C}[1]{>{\centering\let\newline\\\arraybackslash\hspace{0pt}}m{#1}}
\newcolumntype{R}[1]{>{\raggedleft\let\newline\\\arraybackslash\hspace{0pt}}m{#1}}

\begin{document}
\title{Giant radio galaxies in the LOFAR deep fields\textcolor{blue}{$^\star$}}
\author{M. Simonte \inst{1}, H. Andernach \inst{2,3}, M. Br\"uggen \inst{1}, G. K. Miley \inst{4}, P. Barthel \inst{5}}

\institute{Hamburger Sternwarte, University of Hamburg, Gojenbergsweg 112, 21029 Hamburg, Germany; marco.simonte@hs.uni-hamburg.de
\and Th\"uringer Landessternwarte, Sternwarte 5, D-07778 Tautenburg, Germany 
\and Depto. de Astronom\'ia, Univ. de Guanajuato,
Callej\'on de Jalisco s/n, Guanajuato, C.P. 36023, GTO, Mexico; heinz@ugto.mx
\and Sterrewacht Leiden, University of Leiden, 2300 RA, Leiden, The Netherlands
\and Kapteyn Astronomical Institute, University of Groningen, PO Box 800, 9700 AV, Groningen, The Netherlands\\}
%\offprints{%
% E-mail: franco.vazza2@unibo.it}
 %EndAName

 \authorrunning{M. Simonte et al.\inst{1}}% F. Vazza, F. Brighenti, M. Br\"{u}ggen, T. W. Jones}
 \titlerunning{Running title}

\date{Accepted ???. Received ???; in original form ???}

\abstract
  % context heading (optional)
   {The reason why some radio galaxies (RGs) grow to form so-called giant radio galaxies (GRGs) with sizes $>$ 700 kpc, is still unknown.} 
  % aims heading (mandatory)
   {In this study, we compare the radio, optical and environmental properties of GRGs with those of a control sample of smaller RGs we found in the three LOw-Frequency ARray (LOFAR) deep fields, namely the Bo\"otes, ELAIS-N1, Lockman Hole, for a total area of $\approx$ 95 deg$^2$.}
  % methods heading (mandatory)
   {We inspected the LOFAR deep fields and created a catalogue of 1609 extended radio galaxies (ERGs). By visual inspection, we identified their host galaxies and spectroscopically or photometrically classified 280 of these as GRGs. We studied their properties, such as their accretion state, stellar mass and star formation rate (SFR) using deep optical and infrared survey data. Moreover, we explored the environment in terms of the surface number density of neighbouring galaxies within these surveys. Integrated flux densities and radio luminosities were also determined for a subset of ERGs through available survey images at 50, 150, 610, and 1400 MHz to compute integrated spectral indices.}
  % results heading (mandatory)
   {Considering the fraction of GRGs displaying an FRII morphology alongside the host galaxy properties, we suggest that GRGs consistently possess sufficient power to overcome jet frustration caused by the interstellar medium. Moreover, clear differences emerge in the environmental densities between GRGs and smaller RGs, using the number of neighbouring galaxies within 10 Mpc from the host galaxy as a proxy. GRGs preferentially reside in sparser environments compared to their smaller counterparts. In particular, only 3.6\% of the GRGs reside within a 3D comoving distance of 5 Mpc from a previously reported galaxy cluster. We found that larger sources exhibit steeper integrated spectral indices, suggesting that GRGs are late-stage versions of RGs. These results suggest that GRGs are amongst the oldest radio sources with the most stable nuclear activity that reside in sparse environments.}
   {}


%---------------------------------------------------------------------------------------

\maketitle

\section{Introduction}
\label{sec:introduction}

\footnote{\textcolor{blue}{$^\star$} The full catalogue of the 1609 ERGs is only available in electronic form
at the CDS via anonymous ftp to \href{cdsarc.cds.unistra.fr (130.79.128.5)}{cdsarc.cds.unistra.fr (130.79.128.5)}
or via \href{https://cdsarc.cds.unistra.fr/cgi-bin/qcat?J/A+A/}{https://cdsarc.cds.unistra.fr/cgi-bin/qcat?J/A+A/}
}

The study of radio galaxies (RGs) can provide insights into the intricate interplay between active galactic nuclei (AGN), their host galaxies and the surrounding intergalactic medium \citep{Magliocchetti2022}. Among the diverse population of RGs, giant radio galaxies (GRGs) have a linear extent larger than 700 kpc \citep{Willis1974, Barthel1985, Kuzmicz2018}.% and they are relatively rare compared to smaller RGs \citep{Oei2023a}.

The advent of many radio surveys in the past, such as the Faint Images of the Radio Sky at Twenty-cm \citep[FIRST,][]{FIRST1995}, Westerbork Northern Sky Survey \citep[WENSS,][]{WENSS1997}, National Radio Astronomy Observatory (NRAO) VLA Sky Survey \citep[NVSS,][]{NVSS1998}, Sydney University Molonglo Sky Survey \citep[SUMSS,][]{SUMSS2003} and the recent Rapid ASKAP Continuum Survey \citep[RACS,][]{RACS2020} led to the discovery of about 1000 GRGs \citep{Ishwara-Chandra1999, Lara2001, Schoenmakers2000, Machalski2001, Saripalli2005, Kuzmicz2012, Kuzmicz2021, Kuzmicz2018, Dabhade2017, Dabhade2020a, Bruggen2021, Andernach2021, Gurkan2022, Mahato2022}. Furthermore, the LOw-Frequency ARray \citep[LOFAR,][]{LOFAR2013}, with its relatively high resolution and sensitivity to very low surface brightness sources, heralds a new era in the study of very large and high redshift radio galaxies.  The LOFAR Two-metre Sky Survey \citep[LoTSS][]{Shimwell2019, Shimwell2022} has yielded the discovery of approximately 10000 GRGs (\citealt{Dabhade2020b, Simonte2022, Oei2023a}, \citealt{Mostert2023}). Nevertheless, an explanation for the Mpc size of the GRGs is still missing. 

\citet{Oei2023a} carried out a detailed study of the distribution of the linear size of about 2000 GRGs, mainly located at z $<$ 0.4, revealing a relatively steep (power index = $3.5 \pm 0.5$) cumulative power-law distribution. Based on this finding, the authors estimated the number density of GRGs to be $5 \pm 2 (100 \rm ~ Mpc)^{-3}$, emphasising they are truly rare in the local universe. More comprehensive studies of GRGs, extending beyond the local universe, found that GRGs are less rare (\citealt{Simonte2022}, \citealt{Mostert2023}.) than previously thought. \citet{Simonte2022} compared the distribution of the linear size of 74 GRGs with that of a sample of RGs from \citet{Miraghaei2017}. The analysis indicated that the exponential distribution provides an acceptable description of both samples. However, none of these studies could firmly conclude whether GRGs represent the tail of the distribution of the RGs largest linear size.

Models of RGs predict that only very few combinations of jet power, environment, properties of the host galaxy and age of the source can reproduce the large size of GRGs \citep{Turner2015, Hardcastle2018, Turner2023}. The goal of this paper is to constrain the relative abundance of GRGs compared to smaller RGs. Moreover, we study the properties of GRGs and their host galaxies in order to refine models that describe the evolution of RGs. The comparison of the properties of GRGs and smaller RGs can highlight similarities and differences between the two populations. Like smaller RGs, GRGs are hosted by massive elliptical galaxies \citep{Lara2001, Dabhade2017}. The host of the RGs can be classified as High-Excitation (or radiatively efficient) Radio Galaxies (HERG) and Low-Excitation (or radiatively inefficient) Radio Galaxies (LERG) according to their optical spectra \citep{Best2012}. 
%The former have an accretion rate onto the black hole between one and ten per cent of the Eddington rate. HERGs are hosted by bluer, star-forming galaxies and lower-mass black holes. In contrast, LERGs are likely hosted by high-mass galaxies with a central black hole that experiences accretion below one per cent of the Eddington limit \citep{Best2012}. 
LERG is the dominant population, both, for GRGs and smaller RGs \citep{Dabhade2017, Mingo2022, Simonte2022, Best2023}.

The combination of multi-frequency radio observations allows us to study the radio spectrum and thus the radiative age of the emitting electrons in RGs. Previous studies showed that GRGs exhibit radiative ages up to $\approx$ 80 Myr, similar to smaller RGs  \citep{Schoenmakers2000a, Konar2004, Jamrozy2008,Harwood2017,Sebastian2018,Mhlalo2021}. More recent studies used LOFAR to investigate the GRG properties at lower frequencies and derive spectral index maps using data from the literature such as Very Large Array (VLA), Westerbork Synthesis Radio Telescope (WSRT), Effelsberg, and Giant Metrowave Radio Telescope (GMRT) observations \citep{Shulevski2019, Cantwell2020, Dabhade2022}. These studies identified GRGs with a radiative age exceeding 100 Myr, hinting at their status as aged and evolved RGs.

\citet{Lan2021}, compared the statistics of the environment of small and large radio sources. They demonstrated that the properties and the surface number density of surrounding galaxies of GRGs and RGs are very similar. Nevertheless, GRGs are found preferably, but not exclusively, outside of galaxy clusters \citep{Dabhade2020b, Andernach2021} and asymmetries in the morphology of the radio lobes of GRGs appear to be influenced by inhomogeneities of the surrounding medium \citep{Subrahmanyan2008, Safouris2009, Machalski2011, Malarecki2013, Malarecki2015}.

Previous studies mostly focused on the properties of GRGs alone and only very few presented a comparative analysis of the properties of GRGs and smaller RGs \citep[e.g.,][]{Subrahmanyan1996,  Dabhade2020b, Lan2021}. In this paper, we extend our work on the GRGs in the LOFAR Bo\"otes deep field \citep{Simonte2022} by adding a number of RGs and GRGs we found in the Lockman Hole and ELAIS-N1 deep fields. Using radio data for our sources at other frequencies and optical and infrared wavelengths, we compare the multi-wavelength properties of GRGs with those of the control sample of smaller RGs.

The outline of this paper is as follows: in Sec.~\ref{sec:methods} we explain how we built our catalogue of (G)RGs and how we carried out the analyses of the radio and optical data. In Sec.~\ref{sec:results} we present the results of our analysis and compare the properties of RGs and GRGs. We draw our conclusions in Sec.~\ref{sec:conclusions}. Throughout this work we adopt a flat $\rm \Lambda CDM$ cosmology with  \textit{$H_0$} = 70 $\rm ~ km ~ s^{-1} ~ Mpc^{-1}, ~ \Omega_m = 0.3, ~ \Omega_{\Lambda} = 0.7$ and a radio source spectral index $\alpha$ defined as $S_{\nu} \propto \nu^{-\alpha}$.

\section{Methods}
\label{sec:methods}

\begin{figure}[t]
        \centering
        \includegraphics[width= 0.50\textwidth]{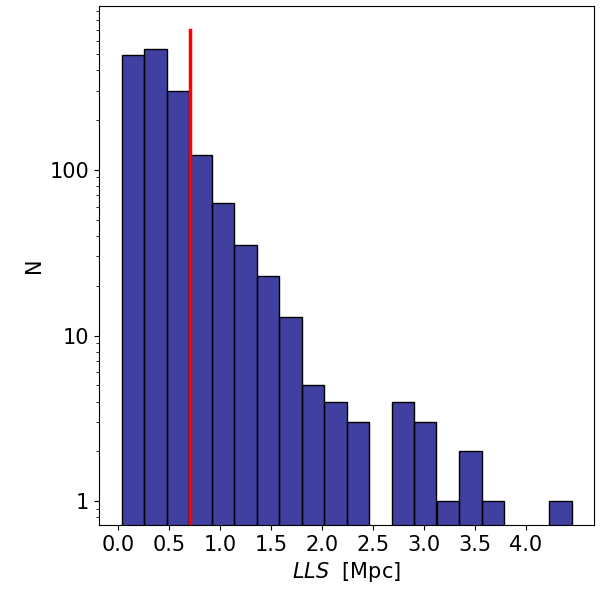}
        \caption{Distribution of the largest linear size in our LDF-RG sample. The red line marks the value of 0.7 Mpc.}
        \label{fig:lls_distribution}
  \end{figure}

\begin{figure}[t!]
        \centering
        \includegraphics[width= 0.50\textwidth]{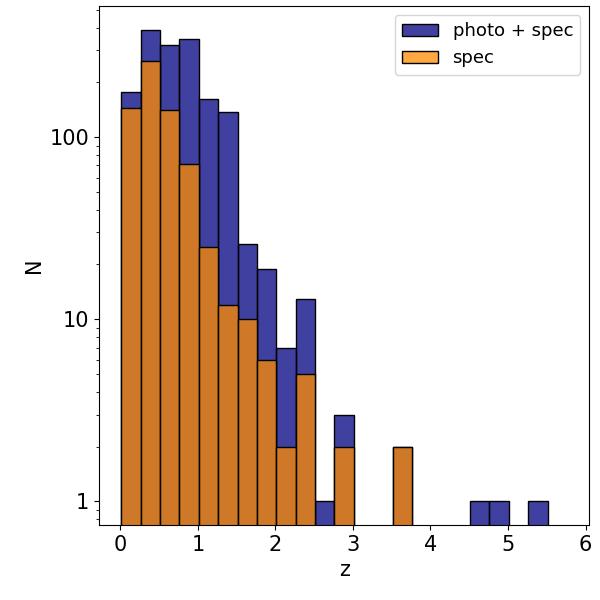}
        \caption{Distribution of the redshifts (spectroscopic and photometric) in our LDF-RG sample (blue histogram). The orange histogram shows the distribution of the spectroscopic redshifts.}
        \label{fig:redshift_distribution}
  \end{figure}

We visually inspected the images of the three LOFAR deep fields: ELAIS-N1, Bo\"otes, which cover an area of $\approx 25 \rm ~deg^2$ each \citep{Sabater2021}, and Lockman Hole, which covers an area of $\approx 45 \rm ~deg^2$ \citep{Tasse2021}. These observations were performed using the LOFAR High-Band Antennas (HBA) which operate in the range between 120-168 MHz. With an effective observing time longer than 80 hours for each field, the observations reach a root mean square noise (rms) level at 150 MHz lower than 30 $\rm \mu Jy ~ beam^{-1}$ across the inner $10 \rm ~deg^2$. The sensitivity to a wide range of angular scales of such observations makes these deep fields ideal for the detection of the diffuse radio emission that characterizes GRGs.

We carried out a systematic search of extended radio galaxies (ERGs) with a largest angular size ($LAS$) $\gtrsim 20^{\prime\prime}$. Given the range of radio morphologies, we employed a variety of methods for estimating the $LAS$. In the case of bright FRII sources, we measured the $LAS$ as the distance between the two opposite hotspots, if detected in the VLA Sky Survey \citep[VLASS,][]{Lacy2020}, unless the LOFAR image showed evidence for emission beyond the hotspots. For FRI and more diffuse sources, we measured the $LAS$ with a straight line between the two opposite ends of the source outlined by the 3-$\sigma$ contours. This approach was also used for bent sources because, otherwise, we would have to make assumptions about the (unknown) 3D structure of the source. For the most diffuse sources, we extended the LAS measurement beyond the 3-sigma contours when we found genuine radio emission beyond those contours.

In order to identify the host galaxies for our ERGs, we made use of a variety of multiple optical and infrared surveys: the Sloan Digital Sky Survey \citep[SDSS,][]{York2000}, the Wide-Field Infrared Survey Explorer \citep[WISE,][]{Wright2010} and its more recent versions of
both images and catalogues such as AllWISE \citep{Cutri2014}, unWISE \citep{Schlafly2019} and CWISE \citep{Marocco2021}, the Dark Energy Survey Imaging \citep[DESI,][]{Dey2019, Zhou2021} and Panoramic Survey Telescope and Rapid Response System \citep[Pan-STARRS,][]{Flewelling2020}. The identification of the host galaxy can be challenging for widely separated radio components without a detected radio core between them. The identification method was extensively explained in previous papers \citep[see][]{Andernach2021, Simonte2022, Simonte2023}. Here, we emphasise the key concerns that may arise during this procedure. 

We first check whether one of the outer radio components has a convincing host itself. In such cases, we recorded the radio source as a separate RG whenever they are larger than $\approx 20^{\prime\prime}$. On the other hand, the presence of a host galaxy with AGN colours \citep{Mateos2012,Assef2013} near the geometrical centre of the ERG is a convincing sign of the genuineness of the ERG. In asymmetric ERGs the shorter lobe is often the brightest, perhaps caused by the denser environment encountered during the jet expansion \citep{Pirya2012, Malarecki2015}, though relativistic effects may also play a role \citep[e.g.,][]{Hocuk2010}. When faced with uncertainty, we relied on this information to locate the host galaxy closer to the brighter lobe for these sources. Moreover, the VLASS images with a resolution of $2.5^{\prime\prime}$ were helpful in spotting the core in a small fraction of ERGs. If neither asymmetry of the source nor the AGN colour were enough to recognise the most likely host galaxies, we chose the brighter or lower redshift host. Hence, the largest linear size (LLS) derived from the LAS and redshift should serve as a lower limit in this case.  For very few sources ($\approx$ 10) we could not find an obvious host either in the optical or in the infrared surveys. Hence, these sources were discarded.

We cross-matched the optical position of the host galaxy with various spectroscopic \citep{Kirshner1987,Wegner2003,Trouille2008,Kochanek2012,Lacy2013,Paris2018,Luo2018, Ahumada2020, Wu2022, Liu-Gebhardt2022, DESIEDR2023} and photometric redshift catalogues \citep{RR2013, Brescia2014, Bilicki2016, Xu2020, Beck2021, Duncan2021, Zhou2021,Wen2021, Duncan2022, Zhang2022, Zou2022}. For the latter, we computed the average redshift and used as associated error the standard deviation of the multiple redshifts of the host galaxy reported in the different catalogues. We do not quote the errors on the spectroscopic redshifts since they are usually more accurate (typical errors are about 0.00015) than the precision we can achieve in the flux density and LAS measurements.

We found 1609 ERGs (which we refer to as LDF-RG or LDF-GRG sample) with an angular size larger than $20^{\prime\prime}$, of which 280 are classified as GRGs (i.e., with an $LLS \ge 0.7$ Mpc) and 134 have an $LLS \ge 1$ Mpc. In total, 42\% of the RGs in our sample have a spectroscopic redshift, 36\% when considering only GRGs. 82 host galaxies do not have a redshift estimate in any of the aforementioned catalogues. 
The \citet{Zhou2021} DESI DR9 photometric redshift catalogue is the deepest full-sky catalogue available (except for the inner regions of the three deep fields where a deeper catalogue from \citealt{Duncan2021} exists). 
In \citet{Zhou2021}, the faintest galaxies have a maximum photometric redshift of around 1.3. Therefore, we assumed a redshift of 1.5 for those host galaxies without redshift listed in the literature. 
However, the precise redshift is not essential when converting LAS to LLS since, for a given LAS, the LLS is almost constant in the redshift range from about 1.2 to 2. (see Fig.~3 in \citealt{Simonte2022}). Moreover, we did not use ERGs without redshift estimates for our environmental analysis. Nevertheless, the uncertainties on the redshift estimation also affect the radio power calculation and thus the location of the ERGs in the radio power-linear size diagram \citep{Baldwin1982}.

We classified the ERGs into FRI and FRII sources according to their radio morphology, without considering the radio luminosity of these sources. In our LDF-GRG sample, 84\% of the GRGs exhibit an FRII morphology, 8\% are classified as FRI, and 24 GRGs are labeled as 'complex' (cpx) due to the ambiguity in their morphology, preventing a robust classification. Considering only the smaller RGs, 74\%, 15\%, 11\% of the ERGs are classified as FRII, FRI and complex, respectively.

In Fig.~\ref{fig:lls_distribution} and \ref{fig:redshift_distribution}, we show the distribution of the LLS and redshift in our LDF-RG sample. These plots encompass all ERGs, including those for which a redshift estimate is not provided in the literature; in such cases, given their faintness we assigned a redshift value of 1.5.

\subsection{Flux density and spectral index}
\label{subsec:flux}

We measured the total radio flux of the 280 GRGs using the full 6$^{\prime\prime}$ resolution images of the LOFAR deep fields \citep{Sabater2021,Tasse2021}. We prepared cutouts with sizes equal to three times the angular size of the source on a side. While integrating the flux, we considered only those pixels whose intensity is larger than $3\sigma_{\rm rms}$. Here, $\sigma_{\rm rms}$ is the noise level calculated within each cutout, since the noise level varies significantly across the deep fields, and it is measured through an iterative approach. In each iteration, we calculated the root mean square (rms) value, removed pixels with intensities exceeding five times the rms, and subsequently recalculated the rms. The convergence criterion was defined based on the difference between two consecutive rms measurements, with a threshold set to 1\%. Therefore, we scaled the noise measurement by the square root of the area of the integrated flux, measured in terms of beam areas. The final flux error was calculated as $\sqrt{\sigma_{\rm rms}^2 + \sigma_{\rm cal}^2}$, where $\sigma_{\rm cal}$ is the uncertainty on the calibration of the flux scale which is assumed to be 10\% \citep{Sabater2021, Tasse2021}. The total flux of some of the largest, in terms of angular size, and faintest ERGs is underestimated with the 3-$\sigma$ clipping method which misses part of the emission coming from the faintest part of the ERG (such as bridges). Thus, we integrated the flux of these 28 sources considering all the pixels belonging to the region of the radio emission.

We calculated the radio power at 150 MHz following: 

\begin{equation}
      P_{150} = 4 \pi D_L^2 S_{150} (1+z)^{\alpha - 1} ,
      \label{eq:power_estimation}
\end{equation}
where $ D_L$ is the luminosity distance, $S_{150}$ is the measured radio flux density at  150 MHz, $(1+z)^{\alpha - 1}$ is the standard k-correction used in radio astronomy and $\alpha$ is the radio spectral index for which we adopted a typical value of 0.7. This is a reasonable value considering that the radio spectrum of GRGs is often dominated by the emission from the hotspots and the core \citep[see,][for a distribution of the integrated spectral indices in a sample of GRGs]{Dabhade2020b}. We should also note that we found slightly steeper radio spectra in our analysis (see LLS-spectra index plots) with an integrated spectral index. Thus, some of the radio luminosities at 150 MHz can be slightly underestimated. Nevertheless, the exact value of $\alpha$ has a relatively small impact on the final estimate of the radio power.

In order to estimate the core fraction, we used a similar approach to calculate the flux of the GRG cores. We only used GRGs in which the core was not blended with the emission coming from the inner part of the jets. We found that 176 GRGs show a clear core emission. 33 GRGs (12\% of the total sample) do not show any core emission above the $3-\sigma$ level. For these sources, we calculated the core flux as 3$\sigma_{rms}$, where $\sigma_{rms}$ is the rms noise calculated in the vicinity of the source. Hence, the core fraction derived from the core flux should serve as an upper limit. The full table with all the properties of the full LDF-RG sample will be made available at the CDS and through the VizieR service\footnote{ \href{https://vizier.cds.unistra.fr}{https://vizier.cds.unistra.fr}} \citep{Ochsenbein2000}. 

Furthermore, we used images at other radio frequencies to investigate the spectral properties of both GRGs and smaller RGs. The LOFAR Low-Band Antenna (LBA) Sky Survey \citep[LoLSS,][]{deGasperin2021, deGasperin2023} is a wide-area survey at 41-66 MHz. A dedicated deep observation of the Bo\"otes field at these frequencies has been performed at a central frequency of 
50 MHz, and the resulting image has an rms noise level of 0.7 mJy beam$^{-1}$ and a resolution of 15$^{\prime\prime}$ \citep{Williams2021}. Moreover, the APERture Tile In Focus array \citep[Apertif,][]{vanCappellen2022} survey, carried out with the Westerbork Synthesis Radio
Telescope (WSRT), observed the same field at 1400 MHz \citep{Kutkin2023}. The mosaicked image has an angular resolution of $27^{\prime\prime}\times11^{\prime\prime}$ and a median background noise of 40 $\mu$Jy beam$^{-1}$. The ELAIS-N1 field was observed with the Giant Metrewave Radio Telescope (GMRT) and imaged at 610 MHz by \citet{Garn2008a}. The resulting radio image has a resolution of $6^{\prime\prime}\times5^{\prime\prime}$ and a rms noise level of 40-60 $\mu$Jy beam$^{-1}$. The Lockman hole field was observed both at 1400 MHz with Apertif \citep{Morganti2021} and 610 MHz with GMRT \citep{Garn2008b}. However, the region of overlap between the LOFAR and the Apertif image in this region of the sky is quite small and the few sources detected at multiple frequencies are too faint to obtain good-quality spectral indices.

Exploiting this multi-frequency coverage we computed the integrated spectral index (i.e., estimated from the total flux of the source) of 25 GRGs and 74 RGs between 150 and 1400 MHz, three GRGs and 32 RGs between 150 and 610 MHz and 24 GRGs and 21 RGs between 50 and 150 MHz. %The radio emission of ERGs is often dominated by emitting electrons residing core and in the hotspots; thus an integrated spectral index (i.e., estimated from the integrated flux of the source) is usually more representative of the emission coming from younger electrons and with a spectral index of about 0.7. On the other hand, computing the spectral index of the source pixel-by-pixel we do not miss the contribution from the faintest and diffuse regions of the ERGs which are likely the locations of the oldest electrons.\\ 
For each pair of images, we convolved the images to the angular resolution of the
image with the coarsest resolution. We excluded those ERGs that were blended with other sources after the convolution. Then, we calculated the integrated flux for each source within the pair of images at two distinct frequencies. This was achieved by applying sigma-clipping at a level of 3$\sigma_{\rm rms}$, where $\sigma_{\rm rms}$ is the rms noise of the individual convolved image. Hence, we calculated the spectral index of the radio spectrum of the source using the linear least-squares method in log-space. The errors were estimated according to the Gaussian propagation of uncertainties. Note that the calculation of the integrated flux is dominated by the brightest part of the radio galaxy, such as the core and the hot spots, which host the youngest electrons and with a spectral index of about 0.4-0.7.

\subsection{Optical analysis}
\label{subsec:opticalanalysis}

\citet{Kondapally2021} and \citet{Best2023} identified the host galaxies for all the radio sources in the inner $\approx 7 \rm ~ deg^2$ of the three LOFAR deep fields and estimated their stellar masses, star formation rates (SFRs) and accretion modes by fitting the spectral energy distribution. We cross-matched our sample of ERG hosts with these catalogues employing a crossmatching radius of 5$^{\prime\prime}$. We found 465 matches, of which 362 have stellar mass and SFR reported in the catalogue. 61 are GRG hosts. We checked the number of ERGs that do not have a match in the \citet{Kondapally2021} and \citet{Best2023} catalogues within a central area of about 9 $\rm deg^2$ of each deep field. We could retrieve 415 of the total 465 matches in the three deep fields. 198 ERGs that are located within the central 9 $\rm deg^2$ of each deep field do not have a host galaxy reported in \citet{Kondapally2021} and \citet{Best2023}. This subset represents RGs that may not be recognized in these catalogues or cases where we chose a different and more likely host galaxy.

The properties of the host galaxies reported in \citet{Kondapally2021} and \citet{Best2023} are derived using the redshift estimates found by \citet{Duncan2021}. In contrast, we determined the redshifts of our sources by averaging the photometric redshifts from multiple catalogues. 
We conducted a comparison between our estimated photometric redshifts with those reported in \citet{Duncan2021}. Our analysis reveals that the redshift difference is consistently below or equal to 0.05 for 75\% of the ERGs and the difference is below 0.2 for 90\% of the host galaxies. While the analysis of the SFR is likely not affected by this discrepancy, it is important to note that this difference affects the discussion on the redshift distribution of GRGs and RGs within this specific analysis.

\subsection{Environmental analysis}
\label{subsec:environmentalanalysis}

We used the DESI DR9 photometric redshift catalogue \citep{Zhou2021} to study the environment of the ERGs on a scale between 1 and 10 Mpc. This catalogue includes observations from the Beijing-Arizona Sky Survey \citep[BASS,][]{Zou2017}, DECam Legacy Survey (DECaLS) and Mayall z-band Legacy Survey (MzLS) \citep{Dey2019} and it is the deepest large-area catalogue available. However, due to the depth limit of the DESI survey we restricted the analysis to $z \leq 0.7$. Hence, we included in our study 810 ERGs of which 121 have a linear size larger than 0.7 Mpc.

First, we created a volume-limited sample of potential neighbouring galaxies. This involves excluding galaxies with an absolute magnitude (luminosity) in the $r$ band that is fainter than the absolute magnitude (luminosity) of a galaxy with an apparent magnitude equal to the flux limit in the $r$ band, observed at the maximum considered redshift ($z=0.7$). Implementing this cut-off imposes a restriction on our analysis as it focuses on the most luminous galaxies, which are also the least common. As a consequence, we might miss faint galaxies residing in galaxy clusters and underestimate the overdensities. 
In this analysis, we looked at the surface number density of galaxies by calculating the number of galaxies within a sphere of radius of 10 Mpc centred on the ERG hosts. We used as photometric redshift of the host galaxy the value obtained by averaging the redshifts of different catalogues as explained in Sec.\ref{sec:methods}. We did not consider the calculated error of the photometric redshift in this analysis. Moreover, we used spectroscopic redshifts for, both, the host and neighbouring galaxies when available. We should point out that more than 99\% of neighbouring galaxies are located at a distance larger than 1 Mpc from the ERG host. Thus, this analysis ignores the information on the environment within $\approx$ 1 Mpc which may play a crucial role in the evolution and expansion of the radio jets and lobes \citep{Konar2008, Subrahmanyan2008, Machalski2011, Pirya2012, Malarecki2015}.

Moreover, we used multiple catalogues of galaxy clusters compiled using observations at different wavelengths \citep{Koester2007,Yoon2008,Hao2010,Rykoff2014, Takey2014, Wen2015, PSZ2, Clerc2016, Burenin2017, Wen2018a, Wen2018b, Abdullah2020, Koulouridis2021} to study the distance of the RGs in our sample from known galaxy clusters, but we restricted these clusters to those with a spectroscopic redshift. We matched the RG hosts with the galaxy cluster having the smallest 3D comoving distance from the host. Due to the sensitivity limitations of the surveys, the detection of galaxy clusters at high redshift is challenging and not many of them are detected beyond $z \approx 0.5$ \citep[e.g][]{Wen2018b}. As a consequence, there may be undetected, high-redshift galaxy clusters closer to the ERG host than the current matched galaxy cluster. Moreover, we may miss the membership of the ERG host to a cluster at high redshift. Thus, we considered only the RGs with a galaxy cluster within a comoving distance of 50 Mpc for the analysis. This value was chosen to study the distribution of the distance of ERGs from galaxy clusters, besides their membership to a specific cluster. The final sample counts 674 RGs and 120 GRGs.

Finally, we looked for a possible cosmological evolution of the linear size of RGs \citep[e.g.,][]{Kapahi1989} by dividing our sample into five redshift bins of width $\Delta z = 0.3$ and calculated the median LLS in the bin. We included all the RGs with a redshift larger than 1.5 in the 1.2-1.5 redshift bin.  We excluded those radio galaxies with an LLS smaller than the minimum physical size required to be resolved at any redshift. We considered an ERG as resolved when its angular size was $\geq 20^{\prime\prime}$ at any redshift. Thus, the minimum LLS required for an ERG to be resolved at any redshift is about 170 kpc. 

\section{Results}
\label{sec:results}

With a sample size of 1609 RGs, we looked for differences in the host galaxy properties of RGs and GRGs and compared the environments in which they reside.

\subsection{$P-D$ diagram}
\label{subsec:PDdiagram}

\begin{figure}[h!]
        \centering
        \includegraphics[width= 0.50\textwidth]{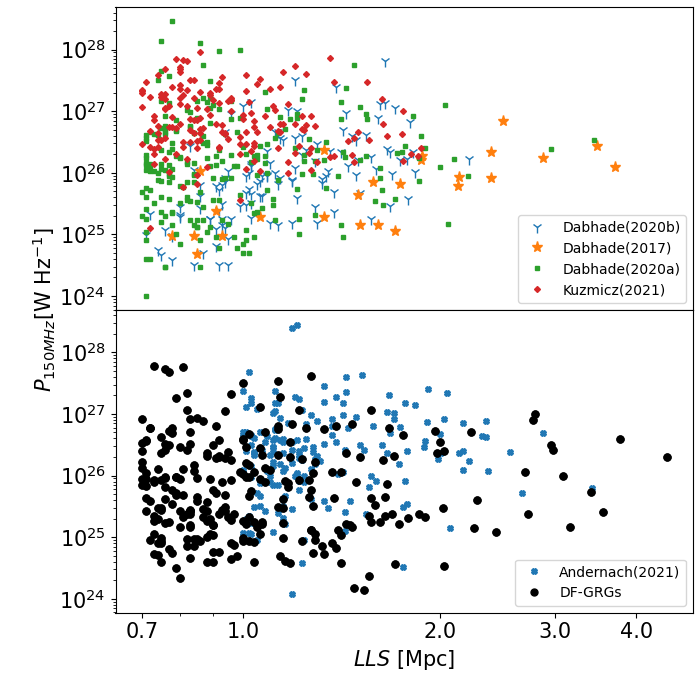}
        \caption{$P-D$ diagram that shows the radio power at 150 MHz against the linear size (LLS) of the LDF-GRG sample. Other previous GRG samples are included in the plot for a comparison: \citet{Dabhade2017, Dabhade2020a, Dabhade2020b, Kuzmicz2021, Andernach2021}.}
        \label{fig:$P-D$_diagram}
  \end{figure}

We used the flux densities and radio powers of the GRGs as derived in Sec.~\ref{subsec:flux} to locate our GRGs in the radio power-linear size diagram \citep[$P-D$ diagram,][]{Baldwin1982}. Such a plot is often used to study the evolution of radio galaxies. Every radio source has a specific evolutionary track in this diagram which depends on multiple properties such as jet power, environment and redshift \citep[see,][]{Ishwara1999, Machalski2004, Hardcastle2018}. We show the position of our LDF-GRGs sample in the $P-D$ diagram with the black points, along with GRGs from recent samples \citep{Dabhade2017, Dabhade2020a, Dabhade2020b, Kuzmicz2021, Andernach2021} in which flux densities and radio power of the listed GRGs are reported. For those GRGs with a flux estimated at other frequencies, we extrapolated the radio power at 150 MHz by using a standard spectral index, $\alpha=0.7$. This is the typical value found in the hotspots of RGs which often dominates the radio emission of the GRGs, especially at higher frequencies (see also Sec.~\ref{subsec:flux}, for a discussion on the value of $\alpha$).

According to the RG models, after an initial phase and once the jet activity stops, its luminosity decreases owing to synchrotron, inverse Compton and adiabatic expansion losses in the latest stages \citep[e.g.,][]{Hardcastle2018}. Hence, if GRGs are aged RGs, it is expected that the largest RGs tend to be less powerful compared to the smaller radio sources. This effect likely drives the lack of GRGs in the upper-right corner of Fig.~\ref{fig:$P-D$_diagram}, where very powerful and large RGs should reside. Figs.~8 and 14 in \citet{Hardcastle2018} show that only the most powerful ($P_{150 MHz} > 10^{26} \rm ~W ~ Hz^{-1}$ and aged ($> 100 \rm ~Myr$) RGs embedded in an environment of $10^{13}-10^{15} \rm ~ M_{\odot}$ can grow up to Mpc sizes. Nevertheless, GRGs display a wide range of radiative ages between 10-100 Myr \citep{Schoenmakers2000, Lara2000, Jamrozy2008, Pinjarkar2023}. Future deep radio observation at higher frequencies will certainly help to accurately estimate the age of GRGs.

On the other hand, there is a lack of very large GRGs with a small radio power as well, likely due to sensitivity limitations of the current radio facilities. The large angular size of such RGs makes their surface brightness rather low, challenging their detection in the lower power regime. As a matter of fact, we have been starting to observe large ($> 1 \rm Mpc$) RGs of low radio luminosity ($< 10^{25} \rm ~W ~ Hz^{-1} $) only with the most recent and sensitive observation with LOFAR and ASKAP \citep{Andernach2021}.

\subsection{Properties of the host galaxies}

\begin{figure}[]
        \centering
        \includegraphics[width= 0.50\textwidth]{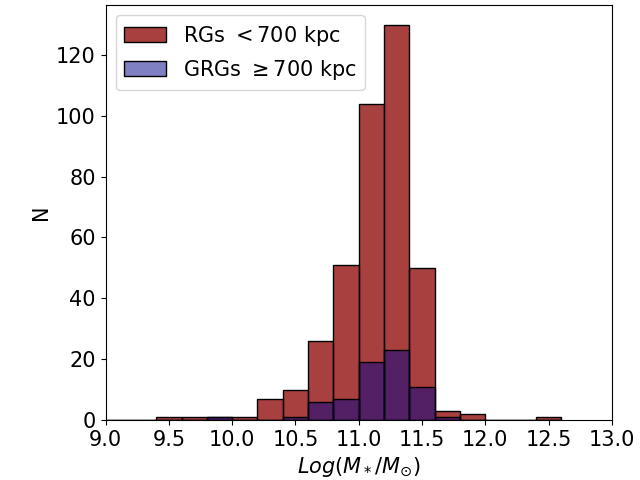}
        \caption{Distribution of the stellar mass of the host galaxies in RGs (red histogram) and GRGs (blue histogram).}
        \label{fig:stellarmass_distirbution}
  \end{figure}

\begin{figure}[]
        \centering
        \includegraphics[width= 0.50\textwidth]{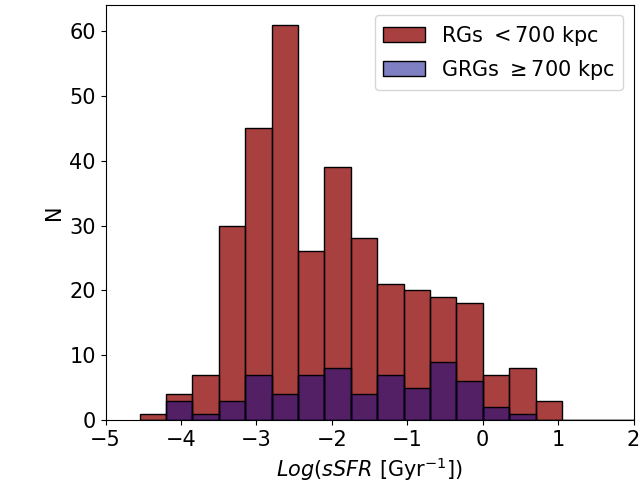}
        \caption{Distribution of the specific star formation rate of the host galaxies in RGs (red histogram) and GRGs (blue histogram).}
        \label{fig:ssfr_distirbution}
  \end{figure}

Combining our LDF-RG sample with the catalogues provided by \citet{Kondapally2021} and \citet{Best2023} for the LOFAR deep fields we compared the star formation rate (SFR), stellar mass and accretion state of the host galaxies in GRGs and RGs. The final sample for this analysis counts 385 RGs ($LLS < 0.7 \rm ~ Mpc$) and 68 GRGs. Fig.~\ref{fig:stellarmass_distirbution} shows that the distribution of the stellar mass in the host of RGs (red) and GRGs (blue) are very similar with both populations being hosted by massive early-type galaxies. The result is in line with previous work \citep{Lara2001, Dabhade2017, Simonte2022}. However, very rare, massive spiral galaxies with radio lobes extending up to Mpc size have been found as well \citep{Bagchi2014, Oei2023b}. Moreover, previous studies \citep{Zovaro2022, Kuzmicz2019} found a young ($<10^7 \rm ~ yr$) and 'intermediate' ($\approx 10^9 \rm ~yr$) stellar population, besides the population of evolved stars with an age larger than $10^{10} \rm ~yr$, in the host of some GRGs. These results suggest that the star formation in some GRG hosts is still ongoing. In our sample, we found that 40\% of GRG hosts have a SFR $> 10 \rm~M_{\odot}~yr^{-1}$,  while only 20\% of the RG hosts have a SFR above this threshold. Interestingly, in our sample, about one-third of the galaxies with a SFR $> 10 \rm~M_{\odot}~yr^{-1}$ host a GRG, while this number drop to 12\% for host galaxies with a SFR $< 10 \rm~M_{\odot}~yr^{-1}$. We compared the SFR distribution of GRGs and smaller RGs by conducting a Kolmogorov-Smirnov (KS) test and we obtained a p-value of 0.004. This result provides a strong evidence to reject the null hypothesis that the two samples were drawn from the same distribution. To validate this result, we performed a Mann-Whitney U test as well \citep{Mann1947}. This is a non-parametric test which uses the rank sums of the two samples to determine whether they come from the same distribution. The resulting p-value of 0.006 further validates the rejection of the null hypothesis. In Fig.~\ref{fig:ssfr_distirbution}, we show the distribution of the specific star formation rate (sSFR) calculated as the ratio between the SFR and the stellar mass of the host galaxy. For comparison, star-forming galaxies are known to span a wide range of sSFR \citep{Gurkan2018} with significant redshift dependence and typical values are larger than 0.1 $\rm Gyr^{-1}$ \citep{Damen2009} across a range extending from the nearby Universe to the median redshift of the sample employed in our analysis, which is 0.7. While the number of RGs decreases at larger sSFR in Fig.~\ref{fig:ssfr_distirbution}, an excess of hosts with moderate and high sSFR ($\gtrsim 0.1 \rm~Gyr^{-1}$) is visible in the sSFR distribution of the GRG hosts with respect to the RG distribution. The returned p-value from both the KS and Mann-Whitney U tests applied to the sSFR distribution is smaller than 0.05, showing that the differences between GRGs and RGs persist even when normalising the SFR by the stellar mass of the galaxy.

While both FRI and FRII RGs are commonly hosted by passive early-type host galaxies, the latter tend to be more star-forming \citep{Mingo2022}. The fraction of FRI sources in the GRG and RG samples used for this analysis is 9\% and 14\%. Moreover, half of the GRG hosts with SFR $> 10 \rm~M_{\odot}~yr^{-1}$ are classified as HERGs, which commonly exhibit larger SFR compared to LERGs. The HERG-LERG classification is taken from \citet{Best2023}. The percentage of HERGs in the GRG and RG samples used for this analysis is slightly different: 22\% and 13\%, respectively.
Thus, the relatively higher presence of HERGs, coupled with the lower fraction of FRI in the GRG sample compared to the RG sample, likely drives the difference in the distribution of the (s)SFR of the two populations. When shifting the linear size threshold used for the definition of GRGs to lower values (e.g., 500 kpc), the percentage of star-forming optical galaxies (SFR $> 10 \rm~M_{\odot}~yr^{-1}$) hosting a GRG increases to 35\%. However, the percentage of optical galaxies with SFR $< 10 \rm~M_{\odot}~yr^{-1}$ hosting a GRG increases to almost 50\% . As a consequence, both, the KS test and the Mann-Whitney U test returns a p-value $> 0.1$. We noticed that the percentage of FRI RGs and HERGs is similar in the GRG and RG samples when employing a threshold of 500 kpc. \citet{Mingo2022} found that the FR morphology of the RG depends on the combination of the radio luminosity of the RG and the stellar mass of the host galaxy. Specifically, a higher galaxy mass or a less powerful radio source increases the likelihood of FRI production. The fraction of FRI in previous GRG samples as well as in the LDF-GRG sample is around 5\% \citep[e.g.,][]{Dabhade2020b, Andernach2021}. Such a low fraction implies that there is a mechanism ensuring that GRGs always possess sufficient power to avoid being frustrated by the host galaxy's environment. It should be noted that in our LDF-GRG sample, there are a few FRIs with a radio luminosity $< 10^{26} \rm ~ W Hz^{-1}$ and stellar masses larger than $10^{11} \rm~M_{\odot}$. However, all these FRI GRGs, for which an estimate of the stellar mass is available, exhibit a bent morphology typical of RGs residing in dense environments, such as galaxy clusters. Within these environments, the lobes of RGs can be advected and stirred by the velocity field of the intracluster medium, resulting in larger and more diffuse lobes (see also Sec.~\ref{subsec:environment_results}). Larger samples of GRGs with available host galaxy mass and SFR are needed to confirm these results.

\begin{figure}
    \centering
    \includegraphics[width= 0.50\textwidth]{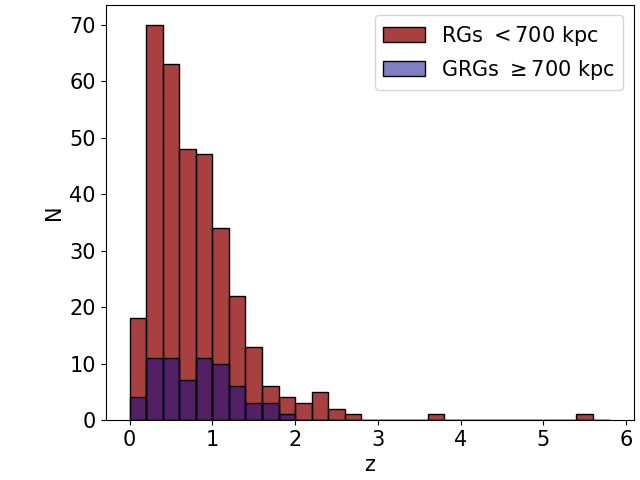}
    \caption{Redshift distributions of the GRG (blue) and RG (red) host galaxies used for the analysis of the (specific) star formation rate.}
    \label{fig:best_redshift_distribution}
\end{figure}

Previous studies have indicated an evolutionary trend in the properties of the RG hosts, revealing that the number density of LERGs exhibits a weak dependence on redshift, whereas HERGs have a higher abundance at higher redshifts \citep{Best2014,Pracy2016, Williams2018, Butler2019, Best2023}. Moreover, a large population of star-forming LERGs has been found at higher redshift \citep{Delvecchio2017, Kondapally2022}. Thus, the different redshift distributions for RGs and GRGs may drive the differences in the distribution of the (s)SFR. We show the redshift distribution of the host galaxies of RGs (red histogram) and GRGs (blue histogram) used for this analysis in Fig.~\ref{fig:best_redshift_distribution}. We examined the similarity between the redshift distributions of the GRG and RG hosts in the present work, for which SFRs are available in \citet{Best2023}. A KS test was performed, yielding a p-value of 0.22. It is worth noticing that while the SFRs reported in \citet{Best2023} are obtained using the redshift catalogue of \citet{Duncan2021}, we averaged the photometric redshifts across multiple catalogues. For this reason, we also performed a KS test and a Mann-Whitney U test on the redshift distributions of GRGs and RGs using the photometric redshift in \citet{Duncan2021}. We found a p-value $>$ 0.05 in both cases.

The tendency of GRGs to exhibit higher (s)SFRs suggests that their host galaxies undergo multiple star-forming phases, possibly resulting in a burst of heightened nuclear activity which can trigger variability in the luminosity of the central AGN \citep[e.g.,][]{Miley1980, Miley2008} or a new jet activity \citep{Gurkan2015, Koziel2017, Shabala2017, Toba2019}. However, we did not find any correlation between the core fraction and the SFR or core radio power and SFR of the host galaxy, possibly due to inherent time delays between star-forming processes and AGN activity. While the star formation observed in a galaxy is contemporaneous with our observations, AGN radio activity likely occurred $10^7-10^8$ years ago. Therefore, establishing a direct link between the radio emission of RGs and the current SFR of their host galaxies is non-trivial, especially considering the uncertainty of whether the present star-forming activity is a result of the same processes that might have triggered AGN activity in the past.
%Alternatively, the continuous fueling of gas triggered by the multiple star-forming phases of the host galaxy may have sustained the jet activity, thus allowing the respective RG to circumvent the quiescent phase.

\begin{figure}[]
        \centering
        \includegraphics[width= 0.50\textwidth]{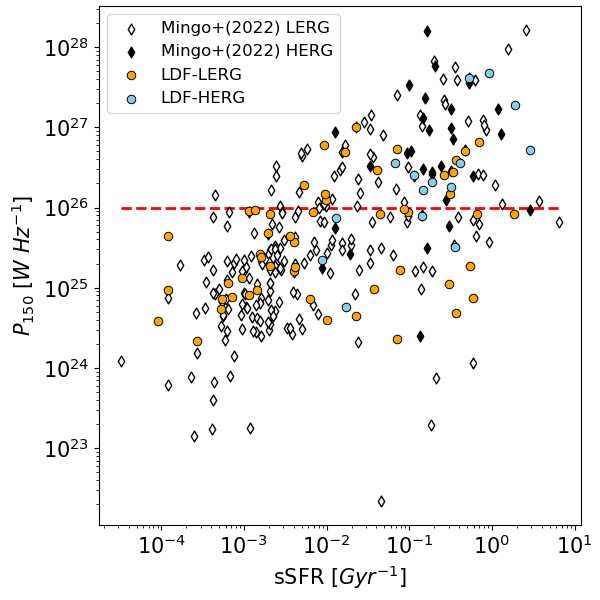}
        \caption{$ \rm sSFR-P_{150MHz}$ diagram for LDF-GRG sample. The orange points are radio galaxy hosts classified as LERGs, while the blue points are HERGs. The red line represents the traditional boundary used to distinguish between FRI and FRII \citep{Fanaroff1974, Ledlow1996}. The filled and empty black diamonds are HERGs and LERGs from the \citet{Mingo2022} sample respectively.}
        \label{fig:ssfr-radiopower_diagram}
  \end{figure}

In Fig.~\ref{fig:ssfr-radiopower_diagram}, we show the sSFR-radio power ($P_{\rm 150MHz}$) diagram for our LDF-GRG sample (orange and blue points) along with the RG sample from \citet{Mingo2022} (black and empty diamonds). The RGs are colour-coded according to their accretion status: black filled diamons and blue points for radiatively efficient and orange and empty diamonds for radiatively inefficient RGs. The threshold value of $10^{26} \rm ~ W ~ Hz^{-1}$ \citep{Ledlow1996, Mingo2022} is represented by the red line.  We notice that RGs with a radiatively efficient accretion have also high radio luminosities ($> 10^{26} \rm ~ W ~ Hz^{-1})$, while they are rare at lower luminosities, as also reported by \citet{Mingo2022}, showing that this dichotomy encompasses ERGs of any size, including GRGs. As expected, LERGs have commonly lower sSFR ($\lesssim 0.1 \rm~Gyr^{-1}$) compared to HERGs. However, there is a population of LERGs with moderate or high sSFR which is driven by a subgroup of FRII sources, both, above and below (but close to) the
traditional luminosity boundary.

It is worth noticing that the apparent relation between the SFR and radio power is not real. As mentioned before, we expect an evolution of the SFR of the host galaxies with cosmic epoch, with a peak of the SFR around z=2 \citep{Madau2014}. 
Furthermore, in brightness-limited surveys such as the LOFAR deep fields, more luminous sources can be observed at a greater redshift. We calculated the partial correlation coefficient \citep{Baba2004} which measures the strength of the correlation between two variables while controlling for the effect of one or more other variables (which is the redshift in our case). The resulting coefficient of R=0.11 suggests that the correlation between the sSFR and the radio power is very mild.

\begin{figure}[]
        \centering
        \includegraphics[width= 0.50\textwidth]{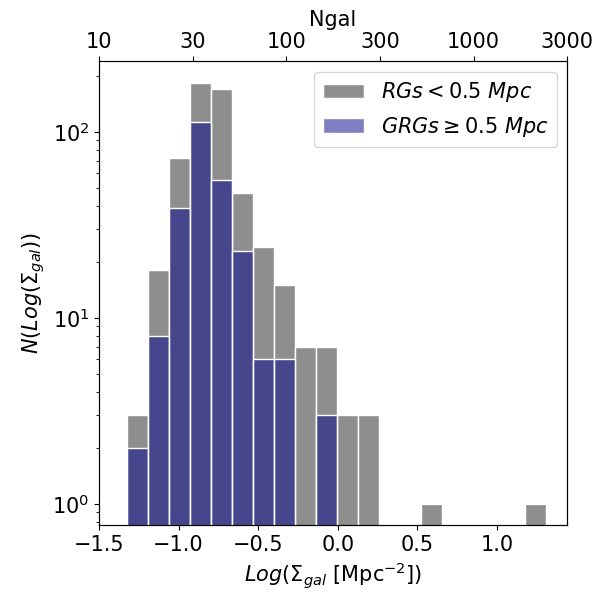}
        \caption{Distribution of the galaxy surface density, within 10 Mpc, around the hosts of RGs smaller (grey) and larger (blue) than 500 kpc. The labels on the top x-axis have the same meaning as those on the lower one, only converted to the number of galaxies within a sphere of 10 Mpc radius.}
        \label{fig:sigma_gal_distribution}
  \end{figure}

\subsection{Environment}
\label{subsec:environment_results}

 We tested the conjecture that GRGs reside in rather sparse environments following the method described in Sec.~\ref{subsec:environmentalanalysis}. In Fig.~\ref{fig:sigma_gal_distribution} we show the distribution of the surface number density of galaxies, $\Sigma_{gal}$, within 10 Mpc. In order to increase the sample of the largest RGs and the significance of our results, we relaxed the definition of GRGs by considering a threshold of 500 kpc rather than the usual 700 kpc. 
 Our sample of 810 RGs used for this analysis consists of 255 RGs with an LLS $>$ 500 kpc with redshift up to $z = 0.7$. The distribution shows that only the smaller RGs reside in the densest environment while only very few GRGs larger than 500 kpc have a number of neighbouring galaxies larger than $\approx$ 200 ($\rm Log(\Sigma_{\rm gal}) \gtrsim -0.2$).

 \begin{figure}[]
        \centering
        \includegraphics[width= 0.50\textwidth]{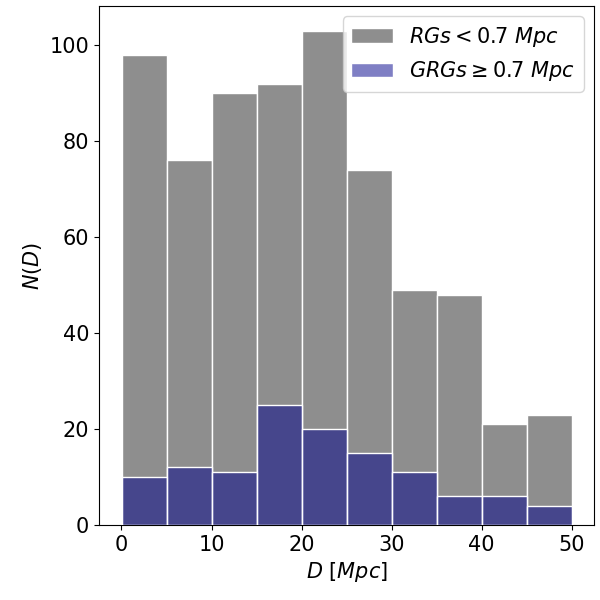}
        \caption{Distribution of the minimum distance of ERG hosts from known galaxy clusters. The blue and grey histograms show the distribution for GRGs ($LLS \ge 700 \rm ~ kpc$) and RGs ($LLS \le 700 \rm ~ kpc$), respectively. }
        \label{fig:galaxy_cluster_distance}
  \end{figure}
 
 We performed a KS test to compare the distribution of the two which resulted in a p-value $<$ 0.001. Thus, we can reject the hypothesis that the two distributions come from the same distribution with a confidence level $>$99\%. The Mann-Whitney U test returns a similar result with a p-value $<$ 0.001. The result of the KS test does not hold when keeping a threshold of 500 kpc for the smaller galaxies while using the widely accepted definition of GRGs (700 kpc). However, this discrepancy is likely a consequence of the reduced sample size (121 GRGs with a linear size larger than 700 kpc) resulting from the more stringent threshold. Furthermore, the Mann-Whitney U test, with a p-value of 0.014, continues to support a difference in the distribution even with these new thresholds. 
 
 Whenever available, we employed spectroscopic redshifts for both the neighbouring and host galaxies to calculate the comoving distances in our analysis. However, the majority of redshifts used in this analysis are photometric. The average error of the latter is 0.09 for the host galaxies and 0.14 for the neighbouring galaxies with a slight tendency to increase with the redshift estimate. To validate our results we performed the same analysis considering only the RGs whose hosts have a spectroscopic redshift (i.e. the uncertainties on the linear size are only due to the uncertainties on the LAS), up to $z=0.7$, while all the neighbouring galaxies were retained, without consideration for the type of redshift. The employed sample counts 364 RGs (LLS$<$500 kpc) and 157 GRGs (LLS$\ge$500) kpc with spectroscopic redshifts. We obtained a p-value of 0.01 for both the KS and the Mann-Whitney U test. Moreover, we performed such an analysis by using the Early Data Release of the Dark Energy Spectroscopic Instrument \citep{DESIEDR2023}, referred to as DESI EDR in what follows, which provides a catalogue of spectroscopic redshifts obtained with the DESI survey. Thus, we only used spectroscopic redshifts for both the sample of RGs and the sample of neighbouring galaxies used to calculate the galaxy density within 10 Mpc. In this catalogue, there is no dedicated column for the optical magnitude of the astrophysical object. Thus, we did not apply any cut in magnitude. DESI EDR does not cover the Lockman Hole, which limits the number of RGs and neighbouring galaxies we can use in the analysis. The retrieved p-values are larger than 0.10. However, the distribution still shows that only RGs ($<$ 700 kpc) reside in environments with a $\rm Log(\Sigma_{\rm gal}) \gtrsim -0.2$.
 
\citet{Oei2022}, using data from the SDSS DR7 \citep{Abazajian2009}, found that "Alcyoneus", the largest GRG published yet, has only five neighbouring galaxies with similar r-band magnitude to its host within 10 Mpc. This result positions Alcyoneus on the far left end of the distribution shown in Fig.~\ref{fig:sigma_gal_distribution}. \citet{Komberg2009}, using SDSS data, demonstrated that GRGs with redshifts up to 0.1 can be found in various environments, ranging from small groups to rich clusters, although they tend to predominately inhabit sparsely populated environments, in agreement with our study. Moreover, in their study, \citet{Lan2021} reported no difference in the environment within a 1 Mpc radius from the host when comparing a sample of GRGs to smaller RGs. It is worth noting that, although our sample of GRGs is of a similar size, our study focused on the galaxies within 10 Mpc, but is less reliable on scales smaller than 1 Mpc as our algorithm likely misses faint satellites around the main galaxies.

Concerning the association of ERGs with catalogued clusters, we show the distribution of the minimum comoving distance (D) of ERG hosts from known clusters with spectroscopic redshift in Fig.~\ref{fig:galaxy_cluster_distance}. While for RGs smaller than 700 kpc the number of host galaxies per bin remains almost constant in the range of 0-25 Mpc, the GRG distribution shows a lack of host galaxies per bin in the range 0-15 Mpc with respect to the range 15-30 Mpc. In particular, the GRG hosts within 15 Mpc are almost half of the GRGs with the nearest galaxy cluster in the range 15-30 Mpc. This indicates that GRGs tend to avoid the densest environments such as galaxy groups and clusters. Nevertheless, the KS test and Mann-Whitney U test reveal that the differences in the two distributions are not very significant, showing p-values between 0.02 to 0.15, dependent on the specific test and size threshold employed for defining GRGs.

We looked at the 109 RGs that fall within a 3D distance of $<$ 5~Mpc from a reported cluster (from references cited in Sec.~\ref{subsec:environmentalanalysis}) and found seven GRGs larger than 1~Mpc plus another three larger than 0.7 Mpc. A closer inspection of these ten revealed that seven of them are in fact the brightest members of their clusters (BrClG in what follows), of which five are wide-angled tailed RGs (WATs) which is a very common radio morphology for BrClGs. The most prominent BrClG in our sample is HB\,13 (GRG J1032+5644), also known as CGCG 290$-$048, the second-largest cluster-associated ERG in our sample (LLS=2.4\,Mpc). It is the BrClG of SDSS--C4~3047 \citep{Linden2007, Abdullah2020} and GalWCat19 cluster 1279 \citep{Abdullah2020}. Interestingly, the cluster-associated GRG with the largest linear size, J1057+5357, and BrClG of WHL~J105727.9+535756 is of FR\,II radio morphology, which is thought to be rare in clusters. However, we find that more than half of the 109 RG hosts within 5~Mpc from clusters in fact coincide (to within 2$''$) with the listed cluster centre and are thus likely to be the BrClG, since these are most often taken as the cluster centre position. Out of 60 such coincidences between ERG host and cluster centre, more than half (36) are of type FRII with a median LLs of 280~kpc, showing that FRII type sources are not that rare among BrClG.

The presence of ten GRGs in our sample that reside within a comoving distance of 5 Mpc from a galaxy cluster challenges the idea that GRGs reside only in sparse environments. A hypothesis is that they reside only in small clusters or groups with a lower central density of the gas. However, using LOFAR and ASKAP data, \citet{Pasini2022} and \citet{Bockmann2023} did not find any relation between the linear size of RGs and the central density of their host galaxy clusters detected within the eROSITA survey \citep{Merloni2012, Predehl2021, Liu2022} suggesting that radio power is more prominent than ambient density in determining the size of the radio galaxy in clusters. However, as emphasised by Fig.~\ref{fig:$P-D$_diagram}, a relation between the size and the radio power in our sample is not obvious. Nevertheless, variations of density and velocity in the intracluster medium may account for the stirring of the plasma injected by the jets and lead to Mpc size WAT and Narrow-Angled Tail (NAT) RGs with complex morphologies \citep{Srivastava2020, Lusetti2024}. Thus, the physical mechanisms responsible for the shape and extent of the diffuse emission of GRGs may be different for those residing in galaxy clusters.

\begin{figure}[]
        \centering
        \includegraphics[width= 0.5\textwidth]{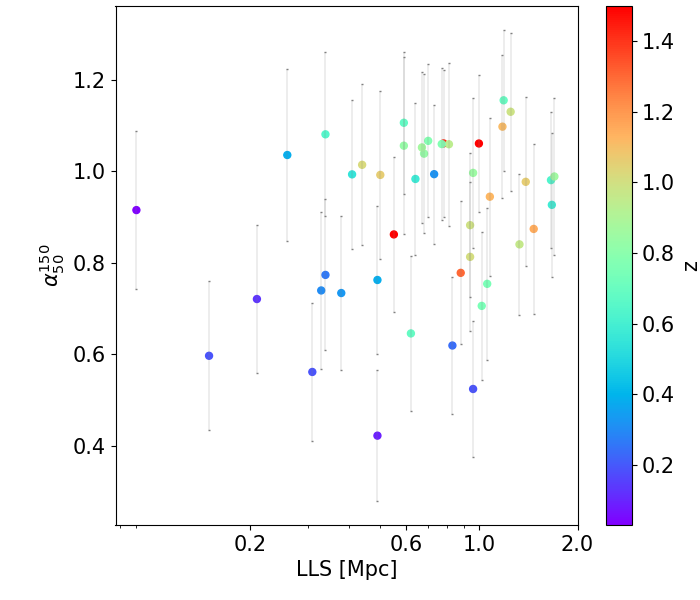}\quad
        \includegraphics[width= 0.5\textwidth]{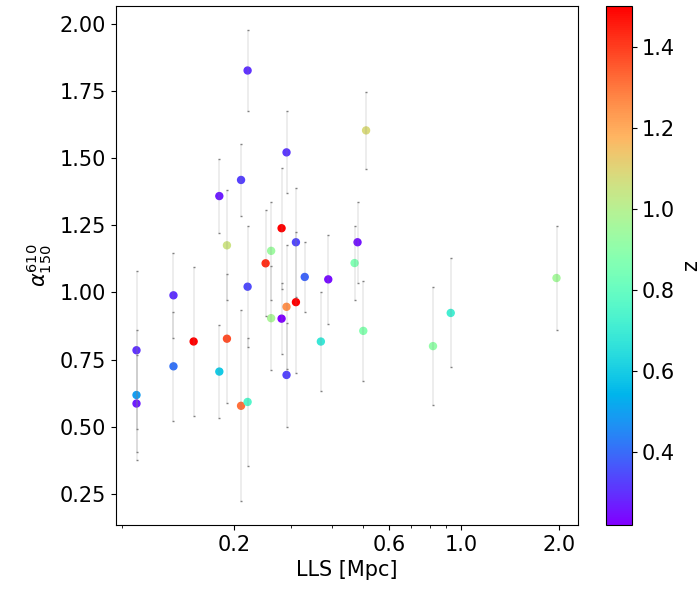}\quad
        \includegraphics[width= 0.5\textwidth]{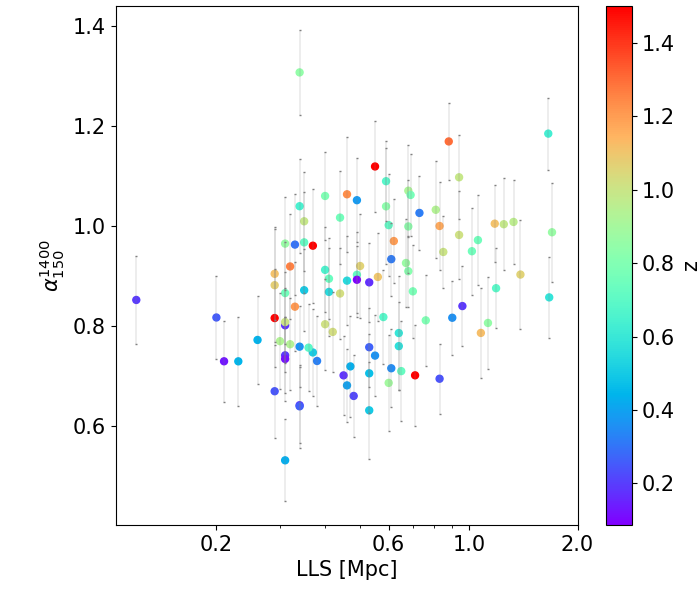}
        \caption{Relation between the integrated spectral index, calculated in the range 50-150 MHz (upper panel), 150-610 MHz (middle panel) and 150-1400 MHz (lower panel), and the largest linear size of RGs. The colorbar highlights the redshift of the sources, while the errorbars represent the error on the spectral index.}
        \label{fig:spidx-lls}
\end{figure}

\subsection{Cosmological size evolution of RGs}
\label{subsec:size_evolution}

Former studies have suggested an evolution of the linear sizes of RGs with redshift, which might be explained by the redshift evolution of the intergalactic medium \citep{Kapahi1989, Machalski2007, Onah2018}. For this analysis, we used both smaller RGs and GRGs, leading to a sample of 1423 ERGs with a linear size larger than 170 kpc. We employed this threshold to remove those ERGs with a linear size smaller than the minimum physical size required to be resolved at any redshift. We found that binning the ERGs in redshift bins of width $\Delta z=0.3$ (see Sec.~\ref{subsec:environment_results}), the median LLS in each bin does not show evolution with redshift. Furthermore, we computed the KS and Mann-Whitney U tests to compare the LLS distribution in different bins of $\Delta z$. We found no evidence that they are drawn from different populations and hence no evidence for cosmological size evolution of RGs. These results are in agreement with \citet{Bruggen2021} who found no dependence of the median linear size or the median radio luminosity on the redshift and hence no evidence for cosmological evolution of the population of GRGs. 

\subsection{Spectral analysis}

According to the hypothesis that GRGs represent a later stage in the evolution of RGs, we expect GRGs to exhibit steeper spectral indices, indicative of the presence of older electrons compared to smaller RGs. We calculated spectral indices for three frequency ranges: 50-150 MHz (LBA, HBA), 150-610 MHz (HBA, GMRT), and 150-1400 MHz (HBA, Apertif). We aimed to identify differences between GRGs and smaller RGs that may indicate the presence of older electrons in the larger RGs. Fig.~\ref{fig:spidx-lls} shows the relation between the linear size of RGs and the integrated spectral index for each frequency pair. The colorbar indicates the redshift of the relevant source and we applied a  redshift limit of z=1.5 in the plots to enhance clarity and visualization. Consequently, the redshift of the sources with the highest values in these plots is equal to or greater than 1.5. However, it is worth noting that a few sources are located at higher redshifts and 13 (G)RGs used in this analysis have a spectroscopic redshift greater or equal to one. When using a combination of Apertif and LOFAR images (lower panel), the two variables show a correlation suggesting that the largest RGs exhibit steeper integrated spectra. We assessed the degree of correlation by calculating the weighted Pearson's correlation coefficient, R. The correlation coefficient ranges between -1 and 1. These limits represent a perfect (anti)correlation, while a value of 0 implies that the variables are uncorrelated. We calculated the weights as the inverse of the squared error of the spectral index. We retrieved R=0.36 corresponding to a p-value $<$ 0.001. Such a result indicates that the largest RGs contain the most aged electrons, pointing to a scenario in which GRGs are in the latest stage of the evolutionary track. It is worth noticing that this correlation is less evident when exclusively considering GRGs ($\rm LLS \gtrsim 700 \rm kpc$), also because of the smaller number of objects in this range. The result may suggest that factors beyond the age of GRGs contribute to the final size of GRGs. The obtained spectral indices between 150 and 1400 MHz are similar to those reported by \citet{Dabhade2020b} who combined LOFAR DR1 and NVSS data. Moreover, previous work suggested a correlation between the size and the radiative age of the RGs \citep{Parma1999, Murgia1999, Jamrozy2008}. Our results support this correlation although we did not directly compute the age of the sources. We also observed a trend involving the spectral index and redshift. It is known that the largest RGs are rarer compared to their smaller counterparts \citep[e.g.,][]{Oei2023a, Simonte2022}; thus, larger cosmological volumes are needed to sample large RGs. Consequently, the correlation between spectral index and redshift is primarily driven by the relationship between linear size and redshift that may appear in an incomplete sample, rather than an independent trend, in our case. Nevertheless, high-redshift RGs have been found originally by their steep spectral index and small angular size \citep{Tielens1979, Blumenthal1979} and the correlation between the spectral index and redshift is still under discussion \citep[e.g.][]{Miley2008}.
The correlation is not significant when using the ERGs detected at 50 MHz (R=0.27,p-value=0.07) and 610 MHz (R=0.03, p-value=0.87) only. However, the lack of correlation might be due to the smaller sample size employed for the analysis.

Using the KS and Mann-Withney U tests, we also analysed the distribution of the integrated spectral index of GRGs and smaller RGs. These tests reject the null hypothesis that the two distributions are drawn from the same population with a confidence level greater or equal to 97\% only when employing a combination of LOFAR HBA-LBA and HBA-Apertif data and not using HBA-GMRT data. The outcomes may indicate that GRGs and their smaller counterparts have a distinct integrated spectral index distribution, even if the results are not conclusive.

\section{Discussion and conclusions}
\label{sec:conclusions}

Even though the number of known GRGs has significantly increased in the past few years (\citealt{Oei2023a}, \citealt{Mostert2023}), the origin and the causes for their large sizes are still not understood. Previous studies mostly focused only on the physics of GRGs themselves without looking for differences between the properties of the largest and smallest ERGs. Here, we visually inspected the three LOFAR deep fields and compiled a catalogue of 1609 ERGs, of which 280 are classified as GRGs. We measured the integrated radio fluxes of the GRGs and located them in the radio power-size (P-D) diagram. The result highlights the ability of LOFAR to detect faint radio emissions from extended RGs. In particular, we reported a few GRGs with a $\rm LLS > 1$~ Mpc and a radio power smaller than $10^{25} \rm ~ W ~ Hz^{-1}$ that were not detected in previous radio surveys. The lack of very powerful ($> 10^{27} \rm ~ W ~ Hz^{-1}$) and large ($\rm LLS > 2 Mpc$) GRGs indicates that the largest GRGs are likely in their latest stage of evolution. Moreover, we used optical catalogues to perform a comprehensive analysis to study the properties of the host galaxies and the environment of GRGs and smaller RGs. To quantify the environmental density, we calculated the galaxy surface density within 10 Mpc from the host galaxy using DESI catalogues \citep{Zhou2021, DESIEDR2023}. We found that the distribution of the galaxy surface density for RGs larger and smaller than 500 kpc is significantly different (p-value=0.002), showing that GRGs statistically reside in sparser environments compared to their smaller counterparts. The result holds when performing the same analysis considering only the ERG hosts with a spectroscopic redshift while retaining all the neighbouring optical galaxies, regardless of their type of redshift. This result does not imply that the environment solely dictates the size of the GRGs. Rather, it indicates that, statistically, within our sample, larger RGs tend to reside in sparser environments compared to their smaller counterparts. Much larger samples with spectroscopic redshift available are needed to firmly conclude that GRGs reside in less densely populated environments.

%Restricting the analysis to only those RGs and neighbouring galaxies for which a spectroscopic redshift is reported in the literature, the KS test returns a p-value= 0.16. However, the sample size, in this case, is three times smaller and much larger samples with spectroscopic redshift available are needed to firmly conclude that GRGs reside in less densely populated environments.

The host galaxies of RGs and GRGs share similar properties. Both populations are hosted by massive early-type galaxies with similar stellar masses and accretion mode (predominantly LERG). Nevertheless, the distribution of the (s)SFR in GRG hosts shows an excess of host galaxies with a $\rm sSFR > 0.1 ~Gyr^{-1}$ compared to the hosts of smaller RGs.

Moreover, we looked at the radio spectra of some of the sources in the LDF-RG sample to find differences in the spectral properties of GRGs and smaller RGs. We calculated the integrated spectral index between multiple pairs of frequencies (50-150 MHz, 150-610 MHz, 150-1400 MHz) and found a positive correlation between the linear size of the ERGs and the steepness of their radio spectrum when using a combination of LOFAR HBA (150 MHz) and Apertif (1400 MHz) observations. On the other hand, the correlation is less evident when considering other frequencies.  This result supports the idea that the largest ERGs are also evolved RGs hosting more aged electrons compared to their smaller counterparts. 
In summary, we propose that the final linear size of RGs is primarily shaped by the combination of jet power and host galaxy's mass, the properties of the large-scale environment, and the age of the source. The power of the jets of GRGs ensures that the final lobes always expand well beyond the host galaxy, leading to a FRII morphology \citep[see][for details]{Mingo2022}. This is suggested by the very low fraction ($\approx 5\%$) of FRI GRGs. Furthermore, the FRI GRGs exhibit a bent morphology suggesting they reside in a dense and dynamical environment that may facilitate their growth to Mpc sizes due to the advection and stirring of their lobes by the turbulent intracluster medium. Fig.~\ref{fig:sigma_gal_distribution} suggests that differences in the density of the environment just outside the host galaxy and up to 10 Mpc can explain the different linear sizes of medium-sized ($\approx$ 200-500 kpc) RGs and GRGs. The more isolated environment in which host galaxies of GRGs are embedded may also explain the differences in the sSFR distribution between GRGs and smaller RGs. The steeper integrated spectral index of GRGs, compared to smaller RGs, implies their advanced evolutionary stage. This underscores the idea that RGs need considerable time to evolve into structures extending beyond Mpc sizes. The full LoTSS coupled with deep optical and infrared surveys, such as DESI, WEAVE \citep{Jin2023} and Euclid \citep{Euclid2022}, holds the potential to provide the necessary confirmation of these results.
 
\begin{acknowledgements}
This research project made use of the following Python packages: APLpy \citep{aplpy2012}, Astropy \citep{astropy2013} and NumPy \citep{numpy2011}. 
MB acknowledges support from the Deutsche Forschungsgemeinschaft under Germany's Excellence Strategy - EXC 2121 "Quantum Universe" - 390833306 and from the BMBF ErUM-Pro grant 05A2023.

LOFAR \citep{LOFAR2013} is the Low Frequency Array designed and constructed by ASTRON. It has observing, data processing, and data storage facilities in several countries, which are owned by various parties (each with their own funding sources), and that are collectively operated by the ILT foundation under a joint scientific policy. The ILT resources have benefited from the following recent major funding sources: CNRS-INSU, Observatoire de Paris and Université d'Orléans, France; BMBF, MIWF-NRW, MPG, Germany; Science Foundation Ireland (SFI), Department of Business, Enterprise and Innovation (DBEI), Ireland; NWO, The Netherlands; The Science and Technology Facilities Council, UK; Ministry of Science and Higher Education, Poland; The Istituto Nazionale di Astrofisica (INAF), Italy.
This research made use of the Dutch national e-infrastructure with support of the SURF Cooperative (e-infra 180169) and the LOFAR e-infra group. The Jülich LOFAR Long Term Archive and the German LOFAR network are both coordinated and operated by the Jülich Supercomputing Centre (JSC), and computing resources on the supercomputer JUWELS at JSC were provided by the Gauss Centre for Supercomputing e.V. (grant CHTB00) through the John von Neumann Institute for Computing (NIC).
This research made use of the University of Hertfordshire high-performance computing facility and the LOFAR-UK computing facility located at the University of Hertfordshire and supported by STFC [ST/P000096/1], and of the Italian LOFAR IT computing infrastructure supported and operated by INAF, and by the Physics Department of Turin university (under an agreement with Consorzio Interuniversitario per la Fisica Spaziale) at the C3S Supercomputing Centre, Italy.
\end{acknowledgements}

\bibliography{marco}

\begin{thebibliography}{165}
\expandafter\ifx\csname natexlab\endcsname\relax\def\natexlab#1{#1}\fi

\bibitem[{{Abazajian} {et~al.}(2009){Abazajian}, {Adelman-McCarthy},
  {Ag{\"u}eros}, {Allam}, {Allende Prieto}, {An}, {Anderson}, {Anderson},
  {Annis}, {Bahcall}, {Bailer-Jones}, {Barentine}, {Bassett}, {Becker},
  {Beers}, {Bell}, {Belokurov}, {Berlind}, {Berman}, {Bernardi}, {Bickerton},
  {Bizyaev}, {Blakeslee}, {Blanton}, {Bochanski}, {Boroski}, {Brewington},
  {Brinchmann}, {Brinkmann}, {Brunner}, {Budav{\'a}ri}, {Carey}, {Carliles},
  {Carr}, {Castander}, {Cinabro}, {Connolly}, {Csabai}, {Cunha}, {Czarapata},
  {Davenport}, {de Haas}, {Dilday}, {Doi}, {Eisenstein}, {Evans}, {Evans},
  {Fan}, {Friedman}, {Frieman}, {Fukugita}, {G{\"a}nsicke}, {Gates},
  {Gillespie}, {Gilmore}, {Gonzalez}, {Gonzalez}, {Grebel}, {Gunn},
  {Gy{\"o}ry}, {Hall}, {Harding}, {Harris}, {Harvanek}, {Hawley}, {Hayes},
  {Heckman}, {Hendry}, {Hennessy}, {Hindsley}, {Hoblitt}, {Hogan}, {Hogg},
  {Holtzman}, {Hyde}, {Ichikawa}, {Ichikawa}, {Im}, {Ivezi{\'c}}, {Jester},
  {Jiang}, {Johnson}, {Jorgensen}, {Juri{\'c}}, {Kent}, {Kessler}, {Kleinman},
  {Knapp}, {Konishi}, {Kron}, {Krzesinski}, {Kuropatkin}, {Lampeitl},
  {Lebedeva}, {Lee}, {Lee}, {French Leger}, {L{\'e}pine}, {Li}, {Lima}, {Lin},
  {Long}, {Loomis}, {Loveday}, {Lupton}, {Magnier}, {Malanushenko},
  {Malanushenko}, {Mandelbaum}, {Margon}, {Marriner}, {Mart{\'\i}nez-Delgado},
  {Matsubara}, {McGehee}, {McKay}, {Meiksin}, {Morrison}, {Mullally}, {Munn},
  {Murphy}, {Nash}, {Nebot}, {Neilsen}, {Newberg}, {Newman}, {Nichol},
  {Nicinski}, {Nieto-Santisteban}, {Nitta}, {Okamura}, {Oravetz}, {Ostriker},
  {Owen}, {Padmanabhan}, {Pan}, {Park}, {Pauls}, {Peoples}, {Percival}, {Pier},
  {Pope}, {Pourbaix}, {Price}, {Purger}, {Quinn}, {Raddick}, {Re Fiorentin},
  {Richards}, {Richmond}, {Riess}, {Rix}, {Rockosi}, {Sako}, {Schlegel},
  {Schneider}, {Scholz}, {Schreiber}, {Schwope}, {Seljak}, {Sesar}, {Sheldon},
  {Shimasaku}, {Sibley}, {Simmons}, {Sivarani}, {Allyn Smith}, {Smith},
  {Smol{\v{c}}i{\'c}}, {Snedden}, {Stebbins}, {Steinmetz}, {Stoughton},
  {Strauss}, {SubbaRao}, {Suto}, {Szalay}, {Szapudi}, {Szkody}, {Tanaka},
  {Tegmark}, {Teodoro}, {Thakar}, {Tremonti}, {Tucker}, {Uomoto}, {Vanden
  Berk}, {Vandenberg}, {Vidrih}, {Vogeley}, {Voges}, {Vogt}, {Wadadekar},
  {Watters}, {Weinberg}, {West}, {White}, {Wilhite}, {Wonders}, {Yanny},
  {Yocum}, {York}, {Zehavi}, {Zibetti}, \& {Zucker}}]{Abazajian2009}
{Abazajian}, K.~N., {Adelman-McCarthy}, J.~K., {Ag{\"u}eros}, M.~A., {et~al.}
  2009, \apjs, 182, 543

\bibitem[{{Abdullah} {et~al.}(2020){Abdullah}, {Wilson}, {Klypin}, {Old},
  {Praton}, \& {Ali}}]{Abdullah2020}
{Abdullah}, M.~H., {Wilson}, G., {Klypin}, A., {et~al.} 2020, \apjs, 246, 2

\bibitem[{{Ahumada} {et~al.}(2020){Ahumada}, {Allende Prieto}, {Almeida},
  {Anders}, {Anderson}, {Andrews}, {Anguiano}, {Arcodia}, {Armengaud},
  {Aubert}, {Avila}, {Avila-Reese}, {Badenes}, {Balland}, {Barger},
  {Barrera-Ballesteros}, {Basu}, {Bautista}, {Beaton}, {Beers}, {Benavides},
  {Bender}, {Bernardi}, {Bershady}, {Beutler}, {Bidin}, {Bird}, {Bizyaev},
  {Blanc}, {Blanton}, {Boquien}, {Borissova}, {Bovy}, {Brandt}, {Brinkmann},
  {Brownstein}, {Bundy}, {Bureau}, {Burgasser}, {Burtin}, {Cano-D{\'\i}az},
  {Capasso}, {Cappellari}, {Carrera}, {Chabanier}, {Chaplin}, {Chapman},
  {Cherinka}, {Chiappini}, {Doohyun Choi}, {Chojnowski}, {Chung}, {Clerc},
  {Coffey}, {Comerford}, {Comparat}, {da Costa}, {Cousinou}, {Covey}, {Crane},
  {Cunha}, {Ilha}, {Dai}, {Damsted}, {Darling}, {Davidson}, {Davies}, {Dawson},
  {De}, {de la Macorra}, {De Lee}, {Queiroz}, {Deconto Machado}, {de la Torre},
  {Dell'Agli}, {du Mas des Bourboux}, {Diamond-Stanic}, {Dillon}, {Donor},
  {Drory}, {Duckworth}, {Dwelly}, {Ebelke}, {Eftekharzadeh}, {Davis Eigenbrot},
  {Elsworth}, {Eracleous}, {Erfanianfar}, {Escoffier}, {Fan}, {Farr},
  {Fern{\'a}ndez-Trincado}, {Feuillet}, {Finoguenov}, {Fofie},
  {Fraser-McKelvie}, {Frinchaboy}, {Fromenteau}, {Fu}, {Galbany}, {Garcia},
  {Garc{\'\i}a-Hern{\'a}ndez}, {Garma Oehmichen}, {Ge}, {Geimba Maia},
  {Geisler}, {Gelfand}, {Goddy}, {Gonzalez-Perez}, {Grabowski}, {Green},
  {Grier}, {Guo}, {Guy}, {Harding}, {Hasselquist}, {Hawken}, {Hayes}, {Hearty},
  {Hekker}, {Hogg}, {Holtzman}, {Horta}, {Hou}, {Hsieh}, {Huber}, {Hunt}, {Ider
  Chitham}, {Imig}, {Jaber}, {Jimenez Angel}, {Johnson}, {Jones},
  {J{\"o}nsson}, {Jullo}, {Kim}, {Kinemuchi}, {Kirkpatrick}, {Kite}, {Klaene},
  {Kneib}, {Kollmeier}, {Kong}, {Kounkel}, {Krishnarao}, {Lacerna}, {Lan},
  {Lane}, {Law}, {Le Goff}, {Leung}, {Lewis}, {Li}, {Lian}, {Lin}, {Long},
  {Longa-Pe{\~n}a}, {Lundgren}, {Lyke}, {Mackereth}, {MacLeod}, {Majewski},
  {Manchado}, {Maraston}, {Martini}, {Masseron}, {Masters}, {Mathur},
  {McDermid}, {Merloni}, {Merrifield}, {M{\'e}sz{\'a}ros}, {Miglio}, {Minniti},
  {Minsley}, {Miyaji}, {Mohammad}, {Mosser}, {Mueller}, {Muna},
  {Mu{\~n}oz-Guti{\'e}rrez}, {Myers}, {Nadathur}, {Nair}, {Nandra}, {Correa do
  Nascimento}, {Nevin}, {Newman}, {Nidever}, {Nitschelm}, {Noterdaeme},
  {O'Connell}, {Olmstead}, {Oravetz}, {Oravetz}, {Osorio}, {Pace}, {Padilla},
  {Palanque-Delabrouille}, {Palicio}, {Pan}, {Pan}, {Parker}, {Paviot},
  {Peirani}, {Ram{\'r}ez}, {Penny}, {Percival}, {Perez-Fournon},
  {P{\'e}rez-R{\`a}fols}, {Petitjean}, {Pieri}, {Pinsonneault}, {Poovelil},
  {Povick}, {Prakash}, {Price-Whelan}, {Raddick}, {Raichoor}, {Ray}, {Rembold},
  {Rezaie}, {Riffel}, {Riffel}, {Rix}, {Robin}, {Roman-Lopes},
  {Rom{\'a}n-Z{\'u}{\~n}iga}, {Rose}, {Ross}, {Rossi}, {Rowlands}, {Rubin},
  {Salvato}, {S{\'a}nchez}, {S{\'a}nchez-Menguiano}, {S{\'a}nchez-Gallego},
  {Sayres}, {Schaefer}, {Schiavon}, {Schimoia}, {Schlafly}, {Schlegel},
  {Schneider}, {Schultheis}, {Schwope}, {Seo}, {Serenelli}, {Shafieloo},
  {Shamsi}, {Shao}, {Shen}, {Shetrone}, {Shirley}, {Silva Aguirre}, {Simon},
  {Skrutskie}, {Slosar}, {Smethurst}, {Sobeck}, {Sodi}, {Souto}, {Stark},
  {Stassun}, {Steinmetz}, {Stello}, {Stermer}, {Storchi-Bergmann},
  {Streblyanska}, {Stringfellow}, {Stutz}, {Su{\'a}rez}, {Sun},
  {Taghizadeh-Popp}, {Talbot}, {Tayar}, {Thakar}, {Theriault}, {Thomas},
  {Thomas}, {Tinker}, {Tojeiro}, {Toledo}, {Tremonti}, {Troup}, {Tuttle},
  {Unda-Sanzana}, {Valentini}, {Vargas-Gonz{\'a}lez}, {Vargas-Maga{\~n}a},
  {V{\'a}zquez-Mata}, {Vivek}, {Wake}, {Wang}, {Weaver}, {Weijmans}, {Wild},
  {Wilson}, {Wilson}, {Wolthuis}, {Wood-Vasey}, {Yan}, {Yang}, {Y{\`e}che},
  {Zamora}, {Zarrouk}, {Zasowski}, {Zhang}, {Zhao}, {Zhao}, {Zheng}, {Zheng},
  {Zhu}, \& {Zou}}]{Ahumada2020}
{Ahumada}, R., {Allende Prieto}, C., {Almeida}, A., {et~al.} 2020, \apjs, 249,
  3

\bibitem[{{Andernach} {et~al.}(2021){Andernach}, {Jim{\'e}nez-Andrade}, \&
  {Willis}}]{Andernach2021}
{Andernach}, H., {Jim{\'e}nez-Andrade}, E.~F., \& {Willis}, A.~G. 2021,
  Galaxies, 9, 99

\bibitem[{{Assef} {et~al.}(2013){Assef}, {Stern}, {Kochanek}, {Blain},
  {Brodwin}, {Brown}, {Donoso}, {Eisenhardt}, {Jannuzi}, {Jarrett}, {Stanford},
  {Tsai}, {Wu}, \& {Yan}}]{Assef2013}
{Assef}, R.~J., {Stern}, D., {Kochanek}, C.~S., {et~al.} 2013, \apj, 772, 26

\bibitem[{{Astropy Collaboration} {et~al.}(2013){Astropy Collaboration},
  {Robitaille}, {Tollerud}, {Greenfield}, {Droettboom}, {Bray}, {Aldcroft},
  {Davis}, {Ginsburg}, {Price-Whelan}, {Kerzendorf}, {Conley}, {Crighton},
  {Barbary}, {Muna}, {Ferguson}, {Grollier}, {Parikh}, {Nair}, {Unther},
  {Deil}, {Woillez}, {Conseil}, {Kramer}, {Turner}, {Singer}, {Fox}, {Weaver},
  {Zabalza}, {Edwards}, {Azalee Bostroem}, {Burke}, {Casey}, {Crawford},
  {Dencheva}, {Ely}, {Jenness}, {Labrie}, {Lim}, {Pierfederici}, {Pontzen},
  {Ptak}, {Refsdal}, {Servillat}, \& {Streicher}}]{astropy2013}
{Astropy Collaboration}, {Robitaille}, T.~P., {Tollerud}, E.~J., {et~al.} 2013,
  \aap, 558, A33

\bibitem[{Baba {et~al.}(2004)Baba, Shibata, \& Sibuya}]{Baba2004}
Baba, K., Shibata, R., \& Sibuya, M. 2004, Australian \& New Zealand Journal of
  Statistics, 46, 657

\bibitem[{{Bagchi} {et~al.}(2014){Bagchi}, {Vivek}, {Vikram}, {Hota}, {Biju},
  {Sirothia}, {Srianand}, {Gopal-Krishna}, \& {Jacob}}]{Bagchi2014}
{Bagchi}, J., {Vivek}, M., {Vikram}, V., {et~al.} 2014, \apj, 788, 174

\bibitem[{{Baldwin}(1982)}]{Baldwin1982}
{Baldwin}, J.~E. 1982, in Extragalactic Radio Sources, ed. D.~S. {Heeschen} \&
  C.~M. {Wade}, Vol.~97, 21--24

\bibitem[{{Barthel} {et~al.}(1985){Barthel}, {Schilizzi}, {Miley}, {Jagers}, \&
  {Strom}}]{Barthel1985}
{Barthel}, P.~D., {Schilizzi}, R.~T., {Miley}, G.~K., {Jagers}, W.~J., \&
  {Strom}, R.~G. 1985, \aap, 148, 243

\bibitem[{{Beck} {et~al.}(2021){Beck}, {Szapudi}, {Flewelling}, {Holmberg},
  {Magnier}, \& {Chambers}}]{Beck2021}
{Beck}, R., {Szapudi}, I., {Flewelling}, H., {et~al.} 2021, \mnras, 500, 1633

\bibitem[{{Becker} {et~al.}(1995){Becker}, {White}, \& {Helfand}}]{FIRST1995}
{Becker}, R.~H., {White}, R.~L., \& {Helfand}, D.~J. 1995, \apj, 450, 559

\bibitem[{{Best} \& {Heckman}(2012)}]{Best2012}
{Best}, P.~N. \& {Heckman}, T.~M. 2012, \mnras, 421, 1569

\bibitem[{{Best} {et~al.}(2014){Best}, {Ker}, {Simpson}, {Rigby}, \&
  {Sabater}}]{Best2014}
{Best}, P.~N., {Ker}, L.~M., {Simpson}, C., {Rigby}, E.~E., \& {Sabater}, J.
  2014, \mnras, 445, 955

\bibitem[{{Best} {et~al.}(2023){Best}, {Kondapally}, {Williams}, {Cochrane},
  {Duncan}, {Hale}, {Haskell}, {Ma{\l}ek}, {McCheyne}, {Smith}, {Wang},
  {Botteon}, {Bonato}, {Bondi}, {Calistro Rivera}, {Gao}, {G{\"u}rkan},
  {Hardcastle}, {Jarvis}, {Mingo}, {Miraghaei}, {Morabito}, {Nisbet},
  {Prandoni}, {R{\"o}ttgering}, {Sabater}, {Shimwell}, {Tasse}, \& {van
  Weeren}}]{Best2023}
{Best}, P.~N., {Kondapally}, R., {Williams}, W.~L., {et~al.} 2023, \mnras, 523,
  1729

\bibitem[{{Bilicki} {et~al.}(2016){Bilicki}, {Peacock}, {Jarrett}, {Cluver},
  {Maddox}, {Brown}, {Taylor}, {Hambly}, {Solarz}, {Holwerda}, {Baldry},
  {Loveday}, {Moffett}, {Hopkins}, {Driver}, {Alpaslan}, \&
  {Bland-Hawthorn}}]{Bilicki2016}
{Bilicki}, M., {Peacock}, J.~A., {Jarrett}, T.~H., {et~al.} 2016, \apjs, 225, 5

\bibitem[{{Blumenthal} \& {Miley}(1979)}]{Blumenthal1979}
{Blumenthal}, G. \& {Miley}, G. 1979, \aap, 80, 13

\bibitem[{{B{\"o}ckmann} {et~al.}(2023){B{\"o}ckmann}, {Br{\"u}ggen},
  {Koribalski}, {Veronica}, {Reiprich}, {Bulbul}, {Bahar}, {Balzer},
  {Comparat}, {Garrel}, {Ghirardini}, {G{\"u}rkan}, {Kluge}, {Leahy},
  {Merloni}, {Liu}, {Ramos-Ceja}, {Salvato}, {Sanders}, {Shabala}, \&
  {Zhang}}]{Bockmann2023}
{B{\"o}ckmann}, K., {Br{\"u}ggen}, M., {Koribalski}, B., {et~al.} 2023, \aap,
  677, A188

\bibitem[{{Brescia} {et~al.}(2014){Brescia}, {Cavuoti}, {Longo}, \& {De
  Stefano}}]{Brescia2014}
{Brescia}, M., {Cavuoti}, S., {Longo}, G., \& {De Stefano}, V. 2014, \aap, 568,
  A126

\bibitem[{{Br{\"u}ggen} {et~al.}(2021){Br{\"u}ggen}, {Reiprich}, {Bulbul},
  {Koribalski}, {Andernach}, {Rudnick}, {Hoang}, {Wilber}, {Duchesne},
  {Veronica}, {Pacaud}, {Hopkins}, {Norris}, {Johnston-Hollitt}, {Brown},
  {Bonafede}, {Brunetti}, {Collier}, {Sanders}, {Vardoulaki}, {Venturi},
  {Kapinska}, \& {Marvil}}]{Bruggen2021}
{Br{\"u}ggen}, M., {Reiprich}, T.~H., {Bulbul}, E., {et~al.} 2021, \aap, 647,
  A3

\bibitem[{{Burenin}(2017)}]{Burenin2017}
{Burenin}, R.~A. 2017, Astronomy Letters, 43, 507

\bibitem[{{Butler} {et~al.}(2019){Butler}, {Huynh}, {Kapi{\'n}ska},
  {Delvecchio}, {Smol{\v{c}}i{\'c}}, {Chiappetti}, {Koulouridis}, \&
  {Pierre}}]{Butler2019}
{Butler}, A., {Huynh}, M., {Kapi{\'n}ska}, A., {et~al.} 2019, \aap, 625, A111

\bibitem[{{Cantwell} {et~al.}(2020){Cantwell}, {Bray}, {Croston}, {Scaife},
  {Mulcahy}, {Best}, {Br{\"u}ggen}, {Brunetti}, {Callingham}, {Clarke},
  {Hardcastle}, {Harwood}, {Heald}, {Heesen}, {Iacobelli}, {Jamrozy},
  {Morganti}, {Orr{\'u}}, {O'Sullivan}, {Riseley}, {R{\"o}ttgering},
  {Shulevski}, {Sridhar}, {Tasse}, \& {Van Eck}}]{Cantwell2020}
{Cantwell}, T.~M., {Bray}, J.~D., {Croston}, J.~H., {et~al.} 2020, \mnras, 495,
  143

\bibitem[{{Clerc} {et~al.}(2016){Clerc}, {Merloni}, {Zhang}, {Finoguenov},
  {Dwelly}, {Nandra}, {Collins}, {Dawson}, {Kneib}, {Rozo}, {Rykoff},
  {Sadibekova}, {Brownstein}, {Lin}, {Ridl}, {Salvato}, {Schwope}, {Steinmetz},
  {Seo}, \& {Tinker}}]{Clerc2016}
{Clerc}, N., {Merloni}, A., {Zhang}, Y.~Y., {et~al.} 2016, \mnras, 463, 4490

\bibitem[{{Condon} {et~al.}(1998){Condon}, {Cotton}, {Greisen}, {Yin},
  {Perley}, {Taylor}, \& {Broderick}}]{NVSS1998}
{Condon}, J.~J., {Cotton}, W.~D., {Greisen}, E.~W., {et~al.} 1998, \aj, 115,
  1693

\bibitem[{{Cutri} {et~al.}(2021){Cutri}, {Wright}, {Conrow}, {Fowler},
  {Eisenhardt}, {Grillmair}, {Kirkpatrick}, {Masci}, {McCallon}, {Wheelock},
  {Fajardo-Acosta}, {Yan}, {Benford}, {Harbut}, {Jarrett}, {Lake}, {Leisawitz},
  {Ressler}, {Stanford}, {Tsai}, {Liu}, {Helou}, {Mainzer}, {Gettngs},
  {Gonzalez}, {Hoffman}, {Marsh}, {Padgett}, {Skrutskie}, {Beck}, {Papin}, \&
  {Wittman}}]{Cutri2014}
{Cutri}, R.~M., {Wright}, E.~L., {Conrow}, T., {et~al.} 2021, VizieR Online
  Data Catalog, II/328

\bibitem[{{Dabhade} {et~al.}(2017){Dabhade}, {Gaikwad}, {Bagchi},
  {Pandey-Pommier}, {Sankhyayan}, \& {Raychaudhury}}]{Dabhade2017}
{Dabhade}, P., {Gaikwad}, M., {Bagchi}, J., {et~al.} 2017, \mnras, 469, 2886

\bibitem[{{Dabhade} {et~al.}(2020{\natexlab{a}}){Dabhade}, {Mahato}, {Bagchi},
  {Saikia}, {Combes}, {Sankhyayan}, {R{\"o}ttgering}, {Ho}, {Gaikwad},
  {Raychaudhury}, {Vaidya}, \& {Guiderdoni}}]{Dabhade2020a}
{Dabhade}, P., {Mahato}, M., {Bagchi}, J., {et~al.} 2020{\natexlab{a}}, \aap,
  642, A153

\bibitem[{{Dabhade} {et~al.}(2020{\natexlab{b}}){Dabhade}, {R{\"o}ttgering},
  {Bagchi}, {Shimwell}, {Hardcastle}, {Sankhyayan}, {Morganti}, {Jamrozy},
  {Shulevski}, \& {Duncan}}]{Dabhade2020b}
{Dabhade}, P., {R{\"o}ttgering}, H.~J.~A., {Bagchi}, J., {et~al.}
  2020{\natexlab{b}}, \aap, 635, A5

\bibitem[{{Dabhade} {et~al.}(2022){Dabhade}, {Shimwell}, {Bagchi}, {Saikia},
  {Combes}, {Gaikwad}, {R{\"o}ttgering}, {Mohapatra}, {Ishwara-Chandra},
  {Intema}, \& {Raychaudhury}}]{Dabhade2022}
{Dabhade}, P., {Shimwell}, T.~W., {Bagchi}, J., {et~al.} 2022, \aap, 668, A64

\bibitem[{{Damen} {et~al.}(2009){Damen}, {Labb{\'e}}, {Franx}, {van Dokkum},
  {Taylor}, \& {Gawiser}}]{Damen2009}
{Damen}, M., {Labb{\'e}}, I., {Franx}, M., {et~al.} 2009, \apj, 690, 937

\bibitem[{{de Gasperin} {et~al.}(2023){de Gasperin}, {Edler}, {Williams},
  {Callingham}, {Asabere}, {Br{\"u}ggen}, {Brunetti}, {Dijkema}, {Hardcastle},
  {Iacobelli}, {Offringa}, {Norden}, {R{\"o}ttgering}, {Shimwell}, {van
  Weeren}, {Tasse}, {Bomans}, {Bonafede}, {Botteon}, {Cassano}, {Chy{\.z}y},
  {Cuciti}, {Emig}, {Kadler}, {Miley}, {Mingo}, {Oei}, {Prandoni}, {Schwarz},
  \& {Zarka}}]{deGasperin2023}
{de Gasperin}, F., {Edler}, H.~W., {Williams}, W.~L., {et~al.} 2023, \aap, 673,
  A165

\bibitem[{{de Gasperin} {et~al.}(2021){de Gasperin}, {Williams}, {Best},
  {Br{\"u}ggen}, {Brunetti}, {Cuciti}, {Dijkema}, {Hardcastle}, {Norden},
  {Offringa}, {Shimwell}, {van Weeren}, {Bomans}, {Bonafede}, {Botteon},
  {Callingham}, {Cassano}, {Chy{\.z}y}, {Emig}, {Edler}, {Haverkorn}, {Heald},
  {Heesen}, {Iacobelli}, {Intema}, {Kadler}, {Ma{\l}ek}, {Mevius}, {Miley},
  {Mingo}, {Morabito}, {Sabater}, {Morganti}, {Orr{\'u}}, {Pizzo}, {Prandoni},
  {Shulevski}, {Tasse}, {Vaccari}, {Zarka}, \&
  {R{\"o}ttgering}}]{deGasperin2021}
{de Gasperin}, F., {Williams}, W.~L., {Best}, P., {et~al.} 2021, \aap, 648,
  A104

\bibitem[{{Delvecchio} {et~al.}(2017){Delvecchio}, {Smol{\v{c}}i{\'c}},
  {Zamorani}, {Lagos}, {Berta}, {Delhaize}, {Baran}, {Alexander}, {Rosario},
  {Gonzalez-Perez}, {Ilbert}, {Lacey}, {Le F{\`e}vre}, {Miettinen}, {Aravena},
  {Bondi}, {Carilli}, {Ciliegi}, {Mooley}, {Novak}, {Schinnerer}, {Capak},
  {Civano}, {Fanidakis}, {Herrera Ruiz}, {Karim}, {Laigle}, {Marchesi},
  {McCracken}, {Middleberg}, {Salvato}, \& {Tasca}}]{Delvecchio2017}
{Delvecchio}, I., {Smol{\v{c}}i{\'c}}, V., {Zamorani}, G., {et~al.} 2017, \aap,
  602, A3

\bibitem[{{DESI Collaboration} {et~al.}(2023){DESI Collaboration}, {Adame},
  {Aguilar}, {Ahlen}, {Alam}, {Aldering}, {Alexander}, {Alfarsy}, {Allende
  Prieto}, {Alvarez}, {Alves}, {Anand}, {Andrade-Oliveira}, {Armengaud},
  {Asorey}, {Avila}, {Aviles}, {Bailey}, {Balaguera-Antol{\'\i}nez},
  {Ballester}, {Baltay}, {Bault}, {Bautista}, {Behera}, {Beltran}, {BenZvi},
  {Beraldo e Silva}, {Bermejo-Climent}, {Berti}, {Besuner}, {Beutler},
  {Bianchi}, {Blake}, {Blum}, {Bolton}, {Brieden}, {Brodzeller}, {Brooks},
  {Brown}, {Buckley-Geer}, {Burtin}, {Cabayol-Garcia}, {Cai}, {Canning},
  {Cardiel-Sas}, {Carnero Rosell}, {Castander}, {Cervantes-Cota}, {Chabanier},
  {Chaussidon}, {Chaves-Montero}, {Chen}, {Chuang}, {Claybaugh}, {Cole},
  {Cooper}, {Cuceu}, {Davis}, {Dawson}, {de Belsunce}, {de la Cruz}, {de la
  Macorra}, {de Mattia}, {Demina}, {Demirbozan}, {DeRose}, {Dey}, {Dey},
  {Dhungana}, {Ding}, {Ding}, {Doel}, {Doshi}, {Douglass}, {Edge},
  {Eftekharzadeh}, {Eisenstein}, {Elliott}, {Escoffier}, {Fagrelius}, {Fan},
  {Fanning}, {Fawcett}, {Ferraro}, {Ereza}, {Flaugher}, {Font-Ribera},
  {Forero-S{\'a}nchez}, {Forero-Romero}, {Frenk}, {G{\"a}nsicke},
  {Garc{\'\i}a}, {Garc{\'\i}a-Bellido}, {Garcia-Quintero}, {Garrison},
  {Gil-Mar{\'\i}n}, {Golden-Marx}, {Gontcho}, {Gonzalez-Morales},
  {Gonzalez-Perez}, {Gordon}, {Graur}, {Green}, {Gruen}, {Guy}, {Hadzhiyska},
  {Hahn}, {Han}, {Hanif}, {Herrera-Alcantar}, {Honscheid}, {Hou}, {Howlett},
  {Huterer}, {Ir{\v{s}}i{\v{c}}}, {Ishak}, {Jacques}, {Jana}, {Jiang},
  {Jimenez}, {Jing}, {Joudaki}, {Jullo}, {Juneau}, {Kizhuprakkat},
  {Kara{\c{c}}ayl{\i}}, {Karim}, {Kehoe}, {Kent}, {Khederlarian}, {Kim},
  {Kirkby}, {Kisner}, {Kitaura}, {Kneib}, {Koposov}, {Kov{\'a}cs}, {Kremin},
  {Krolewski}, {L'Huillier}, {Lambert}, {Lamman}, {Lan}, {Landriau}, {Lang},
  {Lange}, {Lasker}, {Le Guillou}, {Leauthaud}, {Levi}, {Li}, {Linder},
  {Lyons}, {Magneville}, {Manera}, {Manser}, {Margala}, {Martini}, {McDonald},
  {Medina}, {Medina-Varela}, {Meisner}, {Mena-Fern{\'a}ndez}, {Meneses-Rizo},
  {Mezcua}, {Miquel}, {Montero-Camacho}, {Moon}, {Moore}, {Moustakas},
  {Mueller}, {Mundet}, {Mu{\~n}oz-Guti{\'e}rrez}, {Myers}, {Nadathur},
  {Napolitano}, {Neveux}, {Newman}, {Nie}, {Nikutta}, {Niz}, {Norberg},
  {Noriega}, {Paillas}, {Palanque-Delabrouille}, {Palmese}, {Zhiwei},
  {Parkinson}, {Penmetsa}, {Percival}, {P{\'e}rez-Fern{\'a}ndez},
  {P{\'e}rez-R{\`a}fols}, {Pieri}, {Poppett}, {Porredon}, {Pothier}, {Prada},
  {Pucha}, {Raichoor}, {Ram{\'\i}rez-P{\'e}rez}, {Ramirez-Solano},
  {Rashkovetskyi}, {Ravoux}, {Rocher}, {Rockosi}, {Ross}, {Rossi}, {Ruggeri},
  {Ruhlmann-Kleider}, {Sabiu}, {Said}, {Saintonge}, {Samushia}, {Sanchez},
  {Saulder}, {Schaan}, {Schlafly}, {Schlegel}, {Scholte}, {Schubnell}, {Seo},
  {Shafieloo}, {Sharples}, {Sheu}, {Silber}, {Sinigaglia}, {Siudek}, {Slepian},
  {Smith}, {Sprayberry}, {Stephey}, {Su{\'a}rez-P{\'e}rez}, {Sun}, {Tan},
  {Tarl{\'e}}, {Tojeiro}, {Ure{\~n}a-L{\'o}pez}, {Vaisakh}, {Valcin}, {Valdes},
  {Valluri}, {Vargas-Maga{\~n}a}, {Variu}, {Verde}, {Walther}, {Wang}, {Wang},
  {Weaver}, {Weaverdyck}, {Wechsler}, {White}, {Xie}, {Yang}, {Y{\`e}che},
  {Yu}, {Yuan}, {Zhang}, {Zhang}, {Zhao}, {Zheng}, {Zhou}, {Zhou}, {Zou},
  {Zou}, \& {Zu}}]{DESIEDR2023}
{DESI Collaboration}, {Adame}, A.~G., {Aguilar}, J., {et~al.} 2023, arXiv
  e-prints, arXiv:2306.06308

\bibitem[{{Dey} {et~al.}(2019){Dey}, {Schlegel}, {Lang}, {Blum}, {Burleigh},
  {Fan}, {Findlay}, {Finkbeiner}, {Herrera}, {Juneau}, {Landriau}, {Levi},
  {McGreer}, {Meisner}, {Myers}, {Moustakas}, {Nugent}, {Patej}, {Schlafly},
  {Walker}, {Valdes}, {Weaver}, {Y{\`e}che}, {Zou}, {Zhou}, {Abareshi},
  {Abbott}, {Abolfathi}, {Aguilera}, {Alam}, {Allen}, {Alvarez}, {Annis},
  {Ansarinejad}, {Aubert}, {Beechert}, {Bell}, {BenZvi}, {Beutler}, {Bielby},
  {Bolton}, {Brice{\~n}o}, {Buckley-Geer}, {Butler}, {Calamida}, {Carlberg},
  {Carter}, {Casas}, {Castander}, {Choi}, {Comparat}, {Cukanovaite}, {Delubac},
  {DeVries}, {Dey}, {Dhungana}, {Dickinson}, {Ding}, {Donaldson}, {Duan},
  {Duckworth}, {Eftekharzadeh}, {Eisenstein}, {Etourneau}, {Fagrelius},
  {Farihi}, {Fitzpatrick}, {Font-Ribera}, {Fulmer}, {G{\"a}nsicke},
  {Gaztanaga}, {George}, {Gerdes}, {Gontcho}, {Gorgoni}, {Green}, {Guy},
  {Harmer}, {Hernandez}, {Honscheid}, {Huang}, {James}, {Jannuzi}, {Jiang},
  {Joyce}, {Karcher}, {Karkar}, {Kehoe}, {Kneib}, {Kueter-Young}, {Lan},
  {Lauer}, {Le Guillou}, {Le Van Suu}, {Lee}, {Lesser}, {Perreault Levasseur},
  {Li}, {Mann}, {Marshall}, {Mart{\'\i}nez-V{\'a}zquez}, {Martini}, {du Mas des
  Bourboux}, {McManus}, {Meier}, {M{\'e}nard}, {Metcalfe},
  {Mu{\~n}oz-Guti{\'e}rrez}, {Najita}, {Napier}, {Narayan}, {Newman}, {Nie},
  {Nord}, {Norman}, {Olsen}, {Paat}, {Palanque-Delabrouille}, {Peng},
  {Poppett}, {Poremba}, {Prakash}, {Rabinowitz}, {Raichoor}, {Rezaie},
  {Robertson}, {Roe}, {Ross}, {Ross}, {Rudnick}, {Safonova}, {Saha},
  {S{\'a}nchez}, {Savary}, {Schweiker}, {Scott}, {Seo}, {Shan}, {Silva},
  {Slepian}, {Soto}, {Sprayberry}, {Staten}, {Stillman}, {Stupak}, {Summers},
  {Sien Tie}, {Tirado}, {Vargas-Maga{\~n}a}, {Vivas}, {Wechsler}, {Williams},
  {Yang}, {Yang}, {Yapici}, {Zaritsky}, {Zenteno}, {Zhang}, {Zhang}, {Zhou}, \&
  {Zhou}}]{Dey2019}
{Dey}, A., {Schlegel}, D.~J., {Lang}, D., {et~al.} 2019, \aj, 157, 168

\bibitem[{{Duncan}(2022)}]{Duncan2022}
{Duncan}, K.~J. 2022, \mnras, 512, 3662

\bibitem[{{Duncan} {et~al.}(2021){Duncan}, {Kondapally}, {Brown}, {Bonato},
  {Best}, {R{\"o}ttgering}, {Bondi}, {Bowler}, {Cochrane}, {G{\"u}rkan},
  {Hardcastle}, {Jarvis}, {Kunert-Bajraszewska}, {Leslie}, {Ma{\l}ek},
  {Morabito}, {O'Sullivan}, {Prandoni}, {Sabater}, {Shimwell}, {Smith}, {Wang},
  {Wo{\l}owska}, \& {Tasse}}]{Duncan2021}
{Duncan}, K.~J., {Kondapally}, R., {Brown}, M.~J.~I., {et~al.} 2021, \aap, 648,
  A4

\bibitem[{{Euclid Collaboration} {et~al.}(2022){Euclid Collaboration},
  {Scaramella}, {Amiaux}, {Mellier}, {Burigana}, {Carvalho}, {Cuillandre}, {Da
  Silva}, {Derosa}, {Dinis}, {Maiorano}, {Maris}, {Tereno}, {Laureijs},
  {Boenke}, {Buenadicha}, {Dupac}, {Gaspar Venancio}, {G{\'o}mez-{\'A}lvarez},
  {Hoar}, {Lorenzo Alvarez}, {Racca}, {Saavedra-Criado}, {Schwartz}, {Vavrek},
  {Schirmer}, {Aussel}, {Azzollini}, {Cardone}, {Cropper}, {Ealet}, {Garilli},
  {Gillard}, {Granett}, {Guzzo}, {Hoekstra}, {Jahnke}, {Kitching}, {Maciaszek},
  {Meneghetti}, {Miller}, {Nakajima}, {Niemi}, {Pasian}, {Percival},
  {Pottinger}, {Sauvage}, {Scodeggio}, {Wachter}, {Zacchei}, {Aghanim},
  {Amara}, {Auphan}, {Auricchio}, {Awan}, {Balestra}, {Bender}, {Bodendorf},
  {Bonino}, {Branchini}, {Brau-Nogue}, {Brescia}, {Candini}, {Capobianco},
  {Carbone}, {Carlberg}, {Carretero}, {Casas}, {Castander}, {Castellano},
  {Cavuoti}, {Cimatti}, {Cledassou}, {Congedo}, {Conselice}, {Conversi},
  {Copin}, {Corcione}, {Costille}, {Courbin}, {Degaudenzi}, {Douspis},
  {Dubath}, {Duncan}, {Dusini}, {Farrens}, {Ferriol}, {Fosalba}, {Fourmanoit},
  {Frailis}, {Franceschi}, {Franzetti}, {Fumana}, {Gillis}, {Giocoli},
  {Grazian}, {Grupp}, {Haugan}, {Holmes}, {Hormuth}, {Hudelot}, {Kermiche},
  {Kiessling}, {Kilbinger}, {Kohley}, {Kubik}, {K{\"u}mmel}, {Kunz},
  {Kurki-Suonio}, {Lahav}, {Ligori}, {Lilje}, {Lloro}, {Mansutti}, {Marggraf},
  {Markovic}, {Marulli}, {Massey}, {Maurogordato}, {Melchior}, {Merlin},
  {Meylan}, {Mohr}, {Moresco}, {Morin}, {Moscardini}, {Munari}, {Nichol},
  {Padilla}, {Paltani}, {Peacock}, {Pedersen}, {Pettorino}, {Pires}, {Poncet},
  {Popa}, {Pozzetti}, {Raison}, {Rebolo}, {Rhodes}, {Rix}, {Roncarelli},
  {Rossetti}, {Saglia}, {Schneider}, {Schrabback}, {Secroun}, {Seidel},
  {Serrano}, {Sirignano}, {Sirri}, {Skottfelt}, {Stanco}, {Starck},
  {Tallada-Cresp{\'\i}}, {Tavagnacco}, {Taylor}, {Teplitz}, {Toledo-Moreo},
  {Torradeflot}, {Trifoglio}, {Valentijn}, {Valenziano}, {Verdoes Kleijn},
  {Wang}, {Welikala}, {Weller}, {Wetzstein}, {Zamorani}, {Zoubian}, {Andreon},
  {Baldi}, {Bardelli}, {Boucaud}, {Camera}, {Di Ferdinando}, {Fabbian},
  {Farinelli}, {Galeotta}, {Graci{\'a}-Carpio}, {Maino}, {Medinaceli}, {Mei},
  {Neissner}, {Polenta}, {Renzi}, {Romelli}, {Rosset}, {Sureau}, {Tenti},
  {Vassallo}, {Zucca}, {Baccigalupi}, {Balaguera-Antol{\'\i}nez}, {Battaglia},
  {Biviano}, {Borgani}, {Bozzo}, {Cabanac}, {Cappi}, {Casas}, {Castignani},
  {Colodro-Conde}, {Coupon}, {Courtois}, {Cuby}, {de la Torre}, {Desai},
  {Dole}, {Fabricius}, {Farina}, {Ferreira}, {Finelli}, {Flose-Reimberg},
  {Fotopoulou}, {Ganga}, {Gozaliasl}, {Hook}, {Keihanen}, {Kirkpatrick},
  {Liebing}, {Lindholm}, {Mainetti}, {Martinelli}, {Martinet}, {Maturi},
  {McCracken}, {Metcalf}, {Morgante}, {Nightingale}, {Nucita}, {Patrizii},
  {Potter}, {Riccio}, {S{\'a}nchez}, {Sapone}, {Schewtschenko}, {Schultheis},
  {Scottez}, {Teyssier}, {Tutusaus}, {Valiviita}, {Viel}, {Vriend}, \&
  {Whittaker}}]{Euclid2022}
{Euclid Collaboration}, {Scaramella}, R., {Amiaux}, J., {et~al.} 2022, \aap,
  662, A112

\bibitem[{{Fanaroff} \& {Riley}(1974)}]{Fanaroff1974}
{Fanaroff}, B.~L. \& {Riley}, J.~M. 1974, \mnras, 167, 31P

\bibitem[{{Flewelling} {et~al.}(2020){Flewelling}, {Magnier}, {Chambers},
  {Heasley}, {Holmberg}, {Huber}, {Sweeney}, {Waters}, {Calamida}, {Casertano},
  {Chen}, {Farrow}, {Hasinger}, {Henderson}, {Long}, {Metcalfe}, {Narayan},
  {Nieto-Santisteban}, {Norberg}, {Rest}, {Saglia}, {Szalay}, {Thakar},
  {Tonry}, {Valenti}, {Werner}, {White}, {Denneau}, {Draper}, {Hodapp},
  {Jedicke}, {Kaiser}, {Kudritzki}, {Price}, {Wainscoat}, {Chastel}, {McLean},
  {Postman}, \& {Shiao}}]{Flewelling2020}
{Flewelling}, H.~A., {Magnier}, E.~A., {Chambers}, K.~C., {et~al.} 2020, \apjs,
  251, 7

\bibitem[{{Garn} {et~al.}(2008{\natexlab{a}}){Garn}, {Green}, {Riley}, \&
  {Alexander}}]{Garn2008a}
{Garn}, T., {Green}, D.~A., {Riley}, J.~M., \& {Alexander}, P.
  2008{\natexlab{a}}, \mnras, 383, 75

\bibitem[{{Garn} {et~al.}(2008{\natexlab{b}}){Garn}, {Green}, {Riley}, \&
  {Alexander}}]{Garn2008b}
{Garn}, T., {Green}, D.~A., {Riley}, J.~M., \& {Alexander}, P.
  2008{\natexlab{b}}, \mnras, 387, 1037

\bibitem[{{G{\"u}rkan} {et~al.}(2015){G{\"u}rkan}, {Hardcastle}, {Jarvis},
  {Smith}, {Bourne}, {Dunne}, {Maddox}, {Ivison}, \& {Fritz}}]{Gurkan2015}
{G{\"u}rkan}, G., {Hardcastle}, M.~J., {Jarvis}, M.~J., {et~al.} 2015, \mnras,
  452, 3776

\bibitem[{{G{\"u}rkan} {et~al.}(2018){G{\"u}rkan}, {Hardcastle}, {Smith},
  {Best}, {Bourne}, {Calistro-Rivera}, {Heald}, {Jarvis}, {Prandoni},
  {R{\"o}ttgering}, {Sabater}, {Shimwell}, {Tasse}, \& {Williams}}]{Gurkan2018}
{G{\"u}rkan}, G., {Hardcastle}, M.~J., {Smith}, D.~J.~B., {et~al.} 2018,
  \mnras, 475, 3010

\bibitem[{{G{\"u}rkan} {et~al.}(2022){G{\"u}rkan}, {Prandoni}, {O'Brien},
  {Raja}, {Marchetti}, {Vaccari}, {Driver}, {Taylor}, {Franzen}, {Brown},
  {Shabala}, {Andernach}, {Hopkins}, {Norris}, {Leahy}, {Bilicki},
  {Farajollahi}, {Galvin}, {Heald}, {Koribalski}, {An}, \&
  {Warhurst}}]{Gurkan2022}
{G{\"u}rkan}, G., {Prandoni}, I., {O'Brien}, A., {et~al.} 2022, \mnras, 512,
  6104

\bibitem[{{Hao} {et~al.}(2010){Hao}, {McKay}, {Koester}, {Rykoff}, {Rozo},
  {Annis}, {Wechsler}, {Evrard}, {Siegel}, {Becker}, {Busha}, {Gerdes},
  {Johnston}, \& {Sheldon}}]{Hao2010}
{Hao}, J., {McKay}, T.~A., {Koester}, B.~P., {et~al.} 2010, \apjs, 191, 254

\bibitem[{{Hardcastle}(2018)}]{Hardcastle2018}
{Hardcastle}, M.~J. 2018, \mnras, 475, 2768

\bibitem[{{Harwood} {et~al.}(2017){Harwood}, {Hardcastle}, {Morganti},
  {Croston}, {Br{\"u}ggen}, {Brunetti}, {R{\"o}ttgering}, {Shulevski}, \&
  {White}}]{Harwood2017}
{Harwood}, J.~J., {Hardcastle}, M.~J., {Morganti}, R., {et~al.} 2017, \mnras,
  469, 639

\bibitem[{{Hocuk} \& {Barthel}(2010)}]{Hocuk2010}
{Hocuk}, S. \& {Barthel}, P.~D. 2010, \aap, 523, A9

\bibitem[{{Ishwara-Chandra} \&
  {Saikia}(1999{\natexlab{a}})}]{Ishwara-Chandra1999}
{Ishwara-Chandra}, C.~H. \& {Saikia}, D.~J. 1999{\natexlab{a}}, \mnras, 309,
  100

\bibitem[{{Ishwara-Chandra} \& {Saikia}(1999{\natexlab{b}})}]{Ishwara1999}
{Ishwara-Chandra}, C.~H. \& {Saikia}, D.~J. 1999{\natexlab{b}}, \mnras, 309,
  100

\bibitem[{{Jamrozy} {et~al.}(2008){Jamrozy}, {Konar}, {Machalski}, \&
  {Saikia}}]{Jamrozy2008}
{Jamrozy}, M., {Konar}, C., {Machalski}, J., \& {Saikia}, D.~J. 2008, \mnras,
  385, 1286

\bibitem[{{Jin} {et~al.}(2023){Jin}, {Trager}, {Dalton}, {Aguerri}, {Drew},
  {Falc{\'o}n-Barroso}, {G{\"a}nsicke}, {Hill}, {Iovino}, {Pieri}, {Poggianti},
  {Smith}, {Vallenari}, {Abrams}, {Aguado}, {Antoja}, {Arag{\'o}n-Salamanca},
  {Ascasibar}, {Babusiaux}, {Balcells}, {Barrena}, {Battaglia}, {Belokurov},
  {Bensby}, {Bonifacio}, {Bragaglia}, {Carrasco}, {Carrera}, {Cornwell},
  {Dom{\'\i}nguez-Palmero}, {Duncan}, {Famaey}, {Fari{\~n}a}, {Gonzalez},
  {Guest}, {Hatch}, {Hess}, {Hoskin}, {Irwin}, {Knapen}, {Koposov}, {Kuchner},
  {Laigle}, {Lewis}, {Longhetti}, {Lucatello}, {M{\'e}ndez-Abreu}, {Mercurio},
  {Molaeinezhad}, {Mongui{\'o}}, {Morrison}, {Murphy}, {Peralta de Arriba},
  {P{\'e}rez}, {P{\'e}rez-R{\`a}fols}, {Pic{\'o}}, {Raddi}, {Romero-G{\'o}mez},
  {Royer}, {Siebert}, {Seabroke}, {Som}, {Terrett}, {Thomas}, {Wesson},
  {Worley}, {Alfaro}, {Allende Prieto}, {Alonso-Santiago}, {Amos}, {Ashley},
  {Balaguer-N{\'u} nez}, {Balbinot}, {Bellazzini}, {Benn}, {Berlanas},
  {Bernard}, {Best}, {Bettoni}, {Bianco}, {Bishop}, {Blomqvist}, {Boeche},
  {Bolzonella}, {Bonoli}, {Bosma}, {Britavskiy}, {Busarello}, {Caffau},
  {Cantat-Gaudin}, {Castro-Ginard}, {Couto}, {Carbajo-Hijarrubia}, {Carter},
  {Casamiquela}, {Conrado}, {Corcho-Caballero}, {Costantin}, {Deason}, {de
  Burgos}, {De Grandi}, {Di Matteo}, {Dom{\'\i}nguez-G{\'o}mez}, {Dorda},
  {Drake}, {Dutta}, {Erkal}, {Feltzing}, {Ferr{\'e}-Mateu}, {Feuillet},
  {Figueras}, {Fossati}, {Franciosini}, {Frasca}, {Fumagalli}, {Gallazzi},
  {Garc{\'\i}a-Benito}, {Fusillo}, {Gebran}, {Gilbert}, {Gledhill},
  {Gonz{\'a}lez Delgado}, {Greimel}, {Guarcello}, {Guerra}, {Gullieuszik},
  {Haines}, {Hardcastle}, {Harris}, {Haywood}, {Helmi}, {Hernandez}, {Herrero},
  {Hughes}, {Irsic}, {Jablonka}, {Jarvis}, {Jordi}, {Kondapally}, {Kordopatis},
  {Krogager}, {La Barbera}, {Lam}, {Larsen}, {Lemasle}, {Lewis}, {Lhom{\'e}},
  {Lind}, {Lodi}, {Longobardi}, {Lonoce}, {Magrini}, {Ma{\'\i}z Apell{\'a}niz},
  {Marchal}, {Marco}, {Martin}, {Matsuno}, {Maurogordato}, {Merluzzi},
  {Miralda-Escud{\'e}}, {Molinari}, {Monari}, {Morelli}, {Mottram}, {Naylor},
  {Negueruela}, {Onorbe}, {Pancino}, {Peirani}, {Peletier}, {Pozzetti},
  {Rainer}, {Ramos}, {Read}, {Rossi}, {R{\"o}ttgering},
  {Rubi{\~n}o-Mart{\'\i}n}, {Sabater Montes}, {San Juan}, {Sanna}, {Schallig},
  {Schiavon}, {Schultheis}, {Serra}, {Shimwell}, {Sim{\'o}n-D{\'\i}az},
  {Smith}, {Sordo}, {Sorini}, {Soubiran}, {Starkenburg}, {Steele}, {Stott},
  {Stuik}, {Tolstoy}, {Tortora}, {Tsantaki}, {Van der Swaelmen}, {van Weeren},
  {Vergani}, {Verheijen}, {Verro}, {Vink}, {Vioque}, {Walcher}, {Walton},
  {Wegg}, {Weijmans}, {Williams}, {Wilson}, {Wright}, {Xylakis-Dornbusch},
  {Youakim}, {Zibetti}, \& {Zurita}}]{Jin2023}
{Jin}, S., {Trager}, S.~C., {Dalton}, G.~B., {et~al.} 2023, \mnras
  [\eprint[arXiv]{2212.03981}]

\bibitem[{{Kapahi}(1989)}]{Kapahi1989}
{Kapahi}, V.~K. 1989, \aj, 97, 1

\bibitem[{{Kirshner} {et~al.}(1987){Kirshner}, {Oemler}, {Schechter}, \&
  {Shectman}}]{Kirshner1987}
{Kirshner}, R.~P., {Oemler}, Augustus, J., {Schechter}, P.~L., \& {Shectman},
  S.~A. 1987, \apj, 314, 493

\bibitem[{{Kochanek} {et~al.}(2012){Kochanek}, {Eisenstein}, {Cool},
  {Caldwell}, {Assef}, {Jannuzi}, {Jones}, {Murray}, {Forman}, {Dey}, {Brown},
  {Eisenhardt}, {Gonzalez}, {Green}, \& {Stern}}]{Kochanek2012}
{Kochanek}, C.~S., {Eisenstein}, D.~J., {Cool}, R.~J., {et~al.} 2012, \apjs,
  200, 8

\bibitem[{{Koester} {et~al.}(2007){Koester}, {McKay}, {Annis}, {Wechsler},
  {Evrard}, {Bleem}, {Becker}, {Johnston}, {Sheldon}, {Nichol}, {Miller},
  {Scranton}, {Bahcall}, {Barentine}, {Brewington}, {Brinkmann}, {Harvanek},
  {Kleinman}, {Krzesinski}, {Long}, {Nitta}, {Schneider}, {Sneddin}, {Voges},
  \& {York}}]{Koester2007}
{Koester}, B.~P., {McKay}, T.~A., {Annis}, J., {et~al.} 2007, \apj, 660, 239

\bibitem[{{Komberg} \& {Pashchenko}(2009)}]{Komberg2009}
{Komberg}, B.~V. \& {Pashchenko}, I.~N. 2009, Astronomy Reports, 53, 1086

\bibitem[{{Konar} {et~al.}(2008){Konar}, {Jamrozy}, {Saikia}, \&
  {Machalski}}]{Konar2008}
{Konar}, C., {Jamrozy}, M., {Saikia}, D.~J., \& {Machalski}, J. 2008, \mnras,
  383, 525

\bibitem[{{Konar} {et~al.}(2004){Konar}, {Saikia}, {Ishwara-Chandra}, \&
  {Kulkarni}}]{Konar2004}
{Konar}, C., {Saikia}, D.~J., {Ishwara-Chandra}, C.~H., \& {Kulkarni}, V.~K.
  2004, \mnras, 355, 845

\bibitem[{{Kondapally} {et~al.}(2022){Kondapally}, {Best}, {Cochrane},
  {Sabater}, {Duncan}, {Hardcastle}, {Haskell}, {Mingo}, {R{\"o}ttgering},
  {Smith}, {Williams}, {Bonato}, {Calistro Rivera}, {Gao}, {Hale}, {Ma{\l}ek},
  {Miley}, {Prandoni}, \& {Wang}}]{Kondapally2022}
{Kondapally}, R., {Best}, P.~N., {Cochrane}, R.~K., {et~al.} 2022, \mnras, 513,
  3742

\bibitem[{{Kondapally} {et~al.}(2021){Kondapally}, {Best}, {Hardcastle},
  {Nisbet}, {Bonato}, {Sabater}, {Duncan}, {McCheyne}, {Cochrane}, {Bowler},
  {Williams}, {Shimwell}, {Tasse}, {Croston}, {Goyal}, {Jamrozy}, {Jarvis},
  {Mahatma}, {R{\"o}ttgering}, {Smith}, {Wo{\l}owska}, {Bondi}, {Brienza},
  {Brown}, {Br{\"u}ggen}, {Chambers}, {Garrett}, {G{\"u}rkan}, {Huber},
  {Kunert-Bajraszewska}, {Magnier}, {Mingo}, {Mostert},
  {Nikiel-Wroczy{\'n}ski}, {O'Sullivan}, {Paladino}, {Ploeckinger}, {Prandoni},
  {Rosenthal}, {Schwarz}, {Shulevski}, {Wagenveld}, \& {Wang}}]{Kondapally2021}
{Kondapally}, R., {Best}, P.~N., {Hardcastle}, M.~J., {et~al.} 2021, \aap, 648,
  A3

\bibitem[{{Koulouridis} {et~al.}(2021){Koulouridis}, {Clerc}, {Sadibekova},
  {Chira}, {Drigga}, {Faccioli}, {Le F{\`e}vre}, {Garrel}, {Gaynullina},
  {Gkini}, {Kosiba}, {Pacaud}, {Pierre}, {Ridl}, {Tazhenova}, {Adami},
  {Altieri}, {Baguley}, {Cabanac}, {Cucchetti}, {Khalikova}, {Lieu}, {Melin},
  {Molham}, {Ramos-Ceja}, {Soucail}, {Takey}, \&
  {Valtchanov}}]{Koulouridis2021}
{Koulouridis}, E., {Clerc}, N., {Sadibekova}, T., {et~al.} 2021, \aap, 652, A12

\bibitem[{{Kozie{\l}-Wierzbowska} {et~al.}(2017){Kozie{\l}-Wierzbowska}, {Vale
  Asari}, {Stasi{\'n}ska}, {Sikora}, {Goettems}, \&
  {W{\'o}jtowicz}}]{Koziel2017}
{Kozie{\l}-Wierzbowska}, D., {Vale Asari}, N., {Stasi{\'n}ska}, G., {et~al.}
  2017, \apj, 846, 42

\bibitem[{{Kutkin} {et~al.}(2023){Kutkin}, {Oosterloo}, {Morganti}, {Offringa},
  {Adams}, {Adebahr}, {D{\'e}nes}, {Hess}, {van der Hulst}, {de Blok},
  {Bozkurt}, {van Cappellen}, {Gunst}, {Holties}, {van Leeuwen}, {Loose},
  {Oostrum}, {Vohl}, {Wijnholds}, \& {Ziemke}}]{Kutkin2023}
{Kutkin}, A.~M., {Oosterloo}, T.~A., {Morganti}, R., {et~al.} 2023, \aap, 676,
  A37

\bibitem[{{Ku{\'z}micz} {et~al.}(2019){Ku{\'z}micz}, {Czerny}, \&
  {Wildy}}]{Kuzmicz2019}
{Ku{\'z}micz}, A., {Czerny}, B., \& {Wildy}, C. 2019, \aap, 624, A91

\bibitem[{{Ku{\'z}micz} \& {Jamrozy}(2012)}]{Kuzmicz2012}
{Ku{\'z}micz}, A. \& {Jamrozy}, M. 2012, \mnras, 426, 851

\bibitem[{{Ku{\'z}micz} \& {Jamrozy}(2021)}]{Kuzmicz2021}
{Ku{\'z}micz}, A. \& {Jamrozy}, M. 2021, \apjs, 253, 25

\bibitem[{{Ku{\'z}micz} {et~al.}(2018){Ku{\'z}micz}, {Jamrozy}, {Bronarska},
  {Janda-Boczar}, \& {Saikia}}]{Kuzmicz2018}
{Ku{\'z}micz}, A., {Jamrozy}, M., {Bronarska}, K., {Janda-Boczar}, K., \&
  {Saikia}, D.~J. 2018, \apjs, 238, 9

\bibitem[{{Lacy} {et~al.}(2020){Lacy}, {Baum}, {Chandler}, {Chatterjee},
  {Clarke}, {Deustua}, {English}, {Farnes}, {Gaensler}, {Gugliucci},
  {Hallinan}, {Kent}, {Kimball}, {Law}, {Lazio}, {Marvil}, {Mao}, {Medlin},
  {Mooley}, {Murphy}, {Myers}, {Osten}, {Richards}, {Rosolowsky}, {Rudnick},
  {Schinzel}, {Sivakoff}, {Sjouwerman}, {Taylor}, {White}, {Wrobel},
  {Andernach}, {Beasley}, {Berger}, {Bhatnager}, {Birkinshaw}, {Bower},
  {Brandt}, {Brown}, {Burke-Spolaor}, {Butler}, {Comerford}, {Demorest}, {Fu},
  {Giacintucci}, {Golap}, {G{\"u}th}, {Hales}, {Hiriart}, {Hodge}, {Horesh},
  {Ivezi{\'c}}, {Jarvis}, {Kamble}, {Kassim}, {Liu}, {Loinard}, {Lyons},
  {Masters}, {Mezcua}, {Moellenbrock}, {Mroczkowski}, {Nyland}, {O'Dea},
  {O'Sullivan}, {Peters}, {Radford}, {Rao}, {Robnett}, {Salcido}, {Shen},
  {Sobotka}, {Witz}, {Vaccari}, {van Weeren}, {Vargas}, {Williams}, \&
  {Yoon}}]{Lacy2020}
{Lacy}, M., {Baum}, S.~A., {Chandler}, C.~J., {et~al.} 2020, \pasp, 132, 035001

\bibitem[{{Lacy} {et~al.}(2013){Lacy}, {Ridgway}, {Gates}, {Nielsen}, {Petric},
  {Sajina}, {Urrutia}, {Cox Drews}, {Harrison}, {Seymour}, \&
  {Storrie-Lombardi}}]{Lacy2013}
{Lacy}, M., {Ridgway}, S.~E., {Gates}, E.~L., {et~al.} 2013, \apjs, 208, 24

\bibitem[{{Lan} \& {Prochaska}(2021)}]{Lan2021}
{Lan}, T.-W. \& {Prochaska}, X.~J. 2021, \mnras, 502, 5104

\bibitem[{{Lara} {et~al.}(2000){Lara}, {Mack}, {Lacy}, {Klein}, {Cotton},
  {Feretti}, {Giovannini}, \& {Murgia}}]{Lara2000}
{Lara}, L., {Mack}, K.~H., {Lacy}, M., {et~al.} 2000, \aap, 356, 63

\bibitem[{{Lara} {et~al.}(2001){Lara}, {M{\'a}rquez}, {Cotton}, {Feretti},
  {Giovannini}, {Marcaide}, \& {Venturi}}]{Lara2001}
{Lara}, L., {M{\'a}rquez}, I., {Cotton}, W.~D., {et~al.} 2001, \aap, 378, 826

\bibitem[{{Ledlow} \& {Owen}(1996)}]{Ledlow1996}
{Ledlow}, M.~J. \& {Owen}, F.~N. 1996, \aj, 112, 9

\bibitem[{{Liu} {et~al.}(2022{\natexlab{a}}){Liu}, {Bulbul}, {Ghirardini},
  {Liu}, {Klein}, {Clerc}, {{\"O}zsoy}, {Ramos-Ceja}, {Pacaud}, {Comparat},
  {Okabe}, {Bahar}, {Biffi}, {Brunner}, {Br{\"u}ggen}, {Buchner}, {Ider
  Chitham}, {Chiu}, {Dolag}, {Gatuzz}, {Gonzalez}, {Hoang}, {Lamer}, {Merloni},
  {Nandra}, {Oguri}, {Ota}, {Predehl}, {Reiprich}, {Salvato}, {Schrabback},
  {Sanders}, {Seppi}, \& {Thibaud}}]{Liu2022}
{Liu}, A., {Bulbul}, E., {Ghirardini}, V., {et~al.} 2022{\natexlab{a}}, \aap,
  661, A2

\bibitem[{{Liu} {et~al.}(2022{\natexlab{b}}){Liu}, {Gebhardt}, {Cooper},
  {Davis}, {Schneider}, {Ciardullo}, {Farrow}, {Finkelstein}, {Gronwall},
  {Guo}, {Hill}, {House}, {Jeong}, {Jogee}, {Kollatschny}, {Krumpe},
  {Landriau}, {Chavez Ortiz}, {Zhang}, \& {HETDEX
  Collaboration}}]{Liu-Gebhardt2022}
{Liu}, C., {Gebhardt}, K., {Cooper}, E.~M., {et~al.} 2022{\natexlab{b}}, \apjs,
  261, 24

\bibitem[{{Luo} {et~al.}(2018){Luo}, {Zhao}, {Zhao}, \& {et al.}}]{Luo2018}
{Luo}, A.~L., {Zhao}, Y.~H., {Zhao}, G., \& {et al.} 2018, {VizieR Online Data
  Catalog: LAMOST DR4 catalogs (Luo+, 2018)}, VizieR On-line Data Catalog:
  V/153. Originally published in: 2018RAA..in.prep..L

\bibitem[{{Lusetti} {et~al.}(2024){Lusetti}, {de Gasperin}, {Cuciti},
  {Br{\"u}ggen}, {Spinelli}, {Edler}, {Brunetti}, {van Weeren}, {Botteon}, {Di
  Gennaro}, {Cassano}, {Tasse}, \& {Shimwell}}]{Lusetti2024}
{Lusetti}, G., {de Gasperin}, F., {Cuciti}, V., {et~al.} 2024, \mnras, 528, 141

\bibitem[{{Machalski}(2011)}]{Machalski2011}
{Machalski}, J. 2011, \mnras, 413, 2429

\bibitem[{{Machalski} {et~al.}(2004){Machalski}, {Chyzy}, \&
  {Jamrozy}}]{Machalski2004}
{Machalski}, J., {Chyzy}, K.~T., \& {Jamrozy}, M. 2004, \actaa, 54, 391

\bibitem[{{Machalski} {et~al.}(2007){Machalski}, {Chy{\.z}y}, {Stawarz}, \&
  {Kozie{\l}}}]{Machalski2007}
{Machalski}, J., {Chy{\.z}y}, K.~T., {Stawarz}, {\L}., \& {Kozie{\l}}, D. 2007,
  \aap, 462, 43

\bibitem[{{Machalski} {et~al.}(2001){Machalski}, {Jamrozy}, \&
  {Zola}}]{Machalski2001}
{Machalski}, J., {Jamrozy}, M., \& {Zola}, S. 2001, \aap, 371, 445

\bibitem[{{Madau} \& {Dickinson}(2014)}]{Madau2014}
{Madau}, P. \& {Dickinson}, M. 2014, \araa, 52, 415

\bibitem[{{Magliocchetti}(2022)}]{Magliocchetti2022}
{Magliocchetti}, M. 2022, \aapr, 30, 6

\bibitem[{{Mahato} {et~al.}(2022){Mahato}, {Dabhade}, {Saikia}, {Combes},
  {Bagchi}, {Ho}, \& {Raychaudhury}}]{Mahato2022}
{Mahato}, M., {Dabhade}, P., {Saikia}, D.~J., {et~al.} 2022, \aap, 660, A59

\bibitem[{{Malarecki} {et~al.}(2015){Malarecki}, {Jones}, {Saripalli},
  {Staveley-Smith}, \& {Subrahmanyan}}]{Malarecki2015}
{Malarecki}, J.~M., {Jones}, D.~H., {Saripalli}, L., {Staveley-Smith}, L., \&
  {Subrahmanyan}, R. 2015, \mnras, 449, 955

\bibitem[{{Malarecki} {et~al.}(2013){Malarecki}, {Staveley-Smith}, {Saripalli},
  {Subrahmanyan}, {Jones}, {Duffy}, \& {Rioja}}]{Malarecki2013}
{Malarecki}, J.~M., {Staveley-Smith}, L., {Saripalli}, L., {et~al.} 2013,
  \mnras, 432, 200

\bibitem[{{Mann} \& {Whitney}(1947)}]{Mann1947}
{Mann}, H.~B. \& {Whitney}, D.~R. 1947, The Annals of Mathematical Statistics,
  18, 50

\bibitem[{{Marocco} {et~al.}(2021){Marocco}, {Eisenhardt}, {Fowler},
  {Kirkpatrick}, {Meisner}, {Schlafly}, {Stanford}, {Garcia}, {Caselden},
  {Cushing}, {Cutri}, {Faherty}, {Gelino}, {Gonzalez}, {Jarrett}, {Koontz},
  {Mainzer}, {Marchese}, {Mobasher}, {Schlegel}, {Stern}, {Teplitz}, \&
  {Wright}}]{Marocco2021}
{Marocco}, F., {Eisenhardt}, P. R.~M., {Fowler}, J.~W., {et~al.} 2021, \apjs,
  253, 8

\bibitem[{{Mateos} {et~al.}(2012){Mateos}, {Alonso-Herrero}, {Carrera},
  {Blain}, {Watson}, {Barcons}, {Braito}, {Severgnini}, {Donley}, \&
  {Stern}}]{Mateos2012}
{Mateos}, S., {Alonso-Herrero}, A., {Carrera}, F.~J., {et~al.} 2012, \mnras,
  426, 3271

\bibitem[{{Mauch} {et~al.}(2003){Mauch}, {Murphy}, {Buttery}, {Curran},
  {Hunstead}, {Piestrzynski}, {Robertson}, \& {Sadler}}]{SUMSS2003}
{Mauch}, T., {Murphy}, T., {Buttery}, H.~J., {et~al.} 2003, \mnras, 342, 1117

\bibitem[{{McConnell} {et~al.}(2020){McConnell}, {Hale}, {Lenc}, {Banfield},
  {Heald}, {Hotan}, {Leung}, {Moss}, {Murphy}, {O'Brien}, {Pritchard}, {Raja},
  {Sadler}, {Stewart}, {Thomson}, {Whiting}, {Allison}, {Amy}, {Anderson},
  {Ball}, {Bannister}, {Bell}, {Bock}, {Bolton}, {Bunton}, {Chippendale},
  {Collier}, {Cooray}, {Cornwell}, {Diamond}, {Edwards}, {Gupta}, {Hayman},
  {Heywood}, {Jackson}, {Koribalski}, {Lee-Waddell}, {McClure-Griffiths}, {Ng},
  {Norris}, {Phillips}, {Reynolds}, {Roxby}, {Schinckel}, {Shields},
  {Tremblay}, {Tzioumis}, {Voronkov}, \& {Westmeier}}]{RACS2020}
{McConnell}, D., {Hale}, C.~L., {Lenc}, E., {et~al.} 2020, \pasa, 37, e048

\bibitem[{{Merloni} {et~al.}(2012){Merloni}, {Predehl}, {Becker},
  {B{\"o}hringer}, {Boller}, {Brunner}, {Brusa}, {Dennerl}, {Freyberg},
  {Friedrich}, {Georgakakis}, {Haberl}, {Hasinger}, {Meidinger}, {Mohr},
  {Nandra}, {Rau}, {Reiprich}, {Robrade}, {Salvato}, {Santangelo}, {Sasaki},
  {Schwope}, {Wilms}, \& {German eROSITA Consortium}}]{Merloni2012}
{Merloni}, A., {Predehl}, P., {Becker}, W., {et~al.} 2012, arXiv e-prints,
  arXiv:1209.3114

\bibitem[{{Mhlahlo} \& {Jamrozy}(2021)}]{Mhlalo2021}
{Mhlahlo}, N. \& {Jamrozy}, M. 2021, \mnras, 508, 2910

\bibitem[{{Miley}(1980)}]{Miley1980}
{Miley}, G. 1980, \araa, 18, 165

\bibitem[{{Miley} \& {De Breuck}(2008)}]{Miley2008}
{Miley}, G. \& {De Breuck}, C. 2008, \aapr, 15, 67

\bibitem[{{Mingo} {et~al.}(2022){Mingo}, {Croston}, {Best}, {Duncan},
  {Hardcastle}, {Kondapally}, {Prandoni}, {Sabater}, {Shimwell}, {Williams},
  {Baldi}, {Bonato}, {Bondi}, {Dabhade}, {G{\"u}rkan}, {Ineson},
  {Magliocchetti}, {Miley}, {Pierce}, \& {R{\"o}ttgering}}]{Mingo2022}
{Mingo}, B., {Croston}, J.~H., {Best}, P.~N., {et~al.} 2022, \mnras, 511, 3250

\bibitem[{{Miraghaei} \& {Best}(2017)}]{Miraghaei2017}
{Miraghaei}, H. \& {Best}, P.~N. 2017, \mnras, 466, 4346

\bibitem[{{Morganti} {et~al.}(2021){Morganti}, {Oosterloo}, {Brienza},
  {Jurlin}, {Prandoni}, {Orr{\`u}}, {Shabala}, {Adams}, {Adebahr}, {Best},
  {Coolen}, {Damstra}, {de Blok}, {de Gasperin}, {D{\'e}nes}, {Hardcastle},
  {Hess}, {Hut}, {Kondapally}, {Kutkin}, {Loose}, {Lucero}, {Maan}, {Maccagni},
  {Mingo}, {Moss}, {Mostert}, {Norden}, {Oostrum}, {R{\"o}ttgering}, {Ruiter},
  {Shimwell}, {Schulz}, {Vermaas}, {Vohl}, {van der Hulst}, {van Diepen}, {van
  Leeuwen}, \& {Ziemke}}]{Morganti2021}
{Morganti}, R., {Oosterloo}, T.~A., {Brienza}, M., {et~al.} 2021, \aap, 648, A9

\bibitem[{{Mostert} {et~al.}(2023){Mostert}, {Oei}, {Barkus}, {Alegre},
  {Hardcastle}, {Kenneth}, {R{\"o}ttgering}, {van Weeren}, \&
  {Horton}}]{Mostert2023}
{Mostert}, R.~I.~J., {Oei}, M.~S.~S.~L., {Barkus}, B., {et~al.} 2023, \aap,
  subm.

\bibitem[{{Murgia} {et~al.}(1999){Murgia}, {Fanti}, {Fanti}, {Gregorini},
  {Klein}, {Mack}, \& {Vigotti}}]{Murgia1999}
{Murgia}, M., {Fanti}, C., {Fanti}, R., {et~al.} 1999, \aap, 345, 769

\bibitem[{{Ochsenbein} {et~al.}(2000){Ochsenbein}, {Bauer}, \&
  {Marcout}}]{Ochsenbein2000}
{Ochsenbein}, F., {Bauer}, P., \& {Marcout}, J. 2000, \aaps, 143, 23

\bibitem[{{Oei} {et~al.}(2023{\natexlab{a}}){Oei}, {van Weeren}, {Gast},
  {Botteon}, {Hardcastle}, {Dabhade}, {Shimwell}, {R{\"o}ttgering}, \&
  {Drabent}}]{Oei2023a}
{Oei}, M. S.~S.~L., {van Weeren}, R.~J., {Gast}, A. R.~D.~J.~G.~I.~B., {et~al.}
  2023{\natexlab{a}}, \aap, 672, A163

\bibitem[{{Oei} {et~al.}(2022){Oei}, {van Weeren}, {Hardcastle}, {Botteon},
  {Shimwell}, {Dabhade}, {Gast}, {R{\"o}ttgering}, {Br{\"u}ggen}, {Tasse},
  {Williams}, \& {Shulevski}}]{Oei2022}
{Oei}, M. S.~S.~L., {van Weeren}, R.~J., {Hardcastle}, M.~J., {et~al.} 2022,
  \aap, 660, A2

\bibitem[{{Oei} {et~al.}(2023{\natexlab{b}}){Oei}, {van Weeren}, {Hardcastle},
  {Vazza}, {Shimwell}, {Leclercq}, {Br{\"u}ggen}, \&
  {R{\"o}ttgering}}]{Oei2023b}
{Oei}, M. S.~S.~L., {van Weeren}, R.~J., {Hardcastle}, M.~J., {et~al.}
  2023{\natexlab{b}}, \mnras, 518, 240

\bibitem[{{Onah} {et~al.}(2018){Onah}, {Ubachukwu}, {Odo}, \&
  {Onuchukwu}}]{Onah2018}
{Onah}, C.~I., {Ubachukwu}, A.~A., {Odo}, F.~C., \& {Onuchukwu}, C.~C. 2018,
  \rmxaa, 54, 271

\bibitem[{{P{\^a}ris} {et~al.}(2018){P{\^a}ris}, {Petitjean}, {Aubourg},
  {Myers}, {Streblyanska}, {Lyke}, {Anderson}, {Armengaud}, {Bautista},
  {Blanton}, {Blomqvist}, {Brinkmann}, {Brownstein}, {Brandt}, {Burtin},
  {Dawson}, {de la Torre}, {Georgakakis}, {Gil-Mar{\'\i}n}, {Green}, {Hall},
  {Kneib}, {LaMassa}, {Le Goff}, {MacLeod}, {Mariappan}, {McGreer}, {Merloni},
  {Noterdaeme}, {Palanque-Delabrouille}, {Percival}, {Ross}, {Rossi},
  {Schneider}, {Seo}, {Tojeiro}, {Weaver}, {Weijmans}, {Y{\`e}che}, {Zarrouk},
  \& {Zhao}}]{Paris2018}
{P{\^a}ris}, I., {Petitjean}, P., {Aubourg}, {\'E}., {et~al.} 2018, \aap, 613,
  A51

\bibitem[{{Parma} {et~al.}(1999){Parma}, {Murgia}, {Morganti}, {Capetti}, {de
  Ruiter}, \& {Fanti}}]{Parma1999}
{Parma}, P., {Murgia}, M., {Morganti}, R., {et~al.} 1999, \aap, 344, 7

\bibitem[{{Pasini} {et~al.}(2022){Pasini}, {Br{\"u}ggen}, {Hoang},
  {Ghirardini}, {Bulbul}, {Klein}, {Liu}, {Shimwell}, {Hardcastle}, {Williams},
  {Botteon}, {Gastaldello}, {van Weeren}, {Merloni}, {de Gasperin}, {Bahar},
  {Pacaud}, \& {Ramos-Ceja}}]{Pasini2022}
{Pasini}, T., {Br{\"u}ggen}, M., {Hoang}, D.~N., {et~al.} 2022, \aap, 661, A13

\bibitem[{{Pinjarkar} {et~al.}(2023){Pinjarkar}, {Hardcastle}, {Harwood},
  {Lal}, {Hatfield}, {Jarvis}, {Randriamanakoto}, \& {Whittam}}]{Pinjarkar2023}
{Pinjarkar}, S., {Hardcastle}, M.~J., {Harwood}, J.~J., {et~al.} 2023, \mnras,
  523, 620

\bibitem[{{Pirya} {et~al.}(2012){Pirya}, {Saikia}, {Singh}, \&
  {Chandola}}]{Pirya2012}
{Pirya}, A., {Saikia}, D.~J., {Singh}, M., \& {Chandola}, H.~C. 2012, \mnras,
  426, 758

\bibitem[{{Planck Collaboration} {et~al.}(2016){Planck Collaboration}, {Ade},
  {Aghanim}, {Arnaud}, {Ashdown}, {Aumont}, {Baccigalupi}, {Banday},
  {Barreiro}, {Barrena}, {Bartlett}, {Bartolo}, {Battaner}, {Battye},
  {Benabed}, {Beno{\^\i}t}, {Benoit-L{\'e}vy}, {Bernard}, {Bersanelli},
  {Bielewicz}, {Bikmaev}, {B{\"o}hringer}, {Bonaldi}, {Bonavera}, {Bond},
  {Borrill}, {Bouchet}, {Bucher}, {Burenin}, {Burigana}, {Butler}, {Calabrese},
  {Cardoso}, {Carvalho}, {Catalano}, {Challinor}, {Chamballu}, {Chary},
  {Chiang}, {Chon}, {Christensen}, {Clements}, {Colombi}, {Colombo}, {Combet},
  {Comis}, {Couchot}, {Coulais}, {Crill}, {Curto}, {Cuttaia}, {Dahle},
  {Danese}, {Davies}, {Davis}, {de Bernardis}, {de Rosa}, {de Zotti},
  {Delabrouille}, {D{\'e}sert}, {Dickinson}, {Diego}, {Dolag}, {Dole},
  {Donzelli}, {Dor{\'e}}, {Douspis}, {Ducout}, {Dupac}, {Efstathiou},
  {Eisenhardt}, {Elsner}, {En{\ss}lin}, {Eriksen}, {Falgarone}, {Fergusson},
  {Feroz}, {Ferragamo}, {Finelli}, {Forni}, {Frailis}, {Fraisse}, {Franceschi},
  {Frejsel}, {Galeotta}, {Galli}, {Ganga}, {G{\'e}nova-Santos}, {Giard},
  {Giraud-H{\'e}raud}, {Gjerl{\o}w}, {Gonz{\'a}lez-Nuevo}, {G{\'o}rski},
  {Grainge}, {Gratton}, {Gregorio}, {Gruppuso}, {Gudmundsson}, {Hansen},
  {Hanson}, {Harrison}, {Hempel}, {Henrot-Versill{\'e}},
  {Hern{\'a}ndez-Monteagudo}, {Herranz}, {Hildebrandt}, {Hivon}, {Hobson},
  {Holmes}, {Hornstrup}, {Hovest}, {Huffenberger}, {Hurier}, {Jaffe}, {Jaffe},
  {Jin}, {Jones}, {Juvela}, {Keih{\"a}nen}, {Keskitalo}, {Khamitov}, {Kisner},
  {Kneissl}, {Knoche}, {Kunz}, {Kurki-Suonio}, {Lagache}, {Lamarre}, {Lasenby},
  {Lattanzi}, {Lawrence}, {Leonardi}, {Lesgourgues}, {Levrier}, {Liguori},
  {Lilje}, {Linden-V{\o}rnle}, {L{\'o}pez-Caniego}, {Lubin},
  {Mac{\'\i}as-P{\'e}rez}, {Maggio}, {Maino}, {Mak}, {Mandolesi}, {Mangilli},
  {Martin}, {Mart{\'\i}nez-Gonz{\'a}lez}, {Masi}, {Matarrese}, {Mazzotta},
  {McGehee}, {Mei}, {Melchiorri}, {Melin}, {Mendes}, {Mennella}, {Migliaccio},
  {Mitra}, {Miville-Desch{\^e}nes}, {Moneti}, {Montier}, {Morgante},
  {Mortlock}, {Moss}, {Munshi}, {Murphy}, {Naselsky}, {Nastasi}, {Nati},
  {Natoli}, {Netterfield}, {N{\o}rgaard-Nielsen}, {Noviello}, {Novikov},
  {Novikov}, {Olamaie}, {Oxborrow}, {Paci}, {Pagano}, {Pajot}, {Paoletti},
  {Pasian}, {Patanchon}, {Pearson}, {Perdereau}, {Perotto}, {Perrott},
  {Perrotta}, {Pettorino}, {Piacentini}, {Piat}, {Pierpaoli}, {Pietrobon},
  {Plaszczynski}, {Pointecouteau}, {Polenta}, {Pratt}, {Pr{\'e}zeau}, {Prunet},
  {Puget}, {Rachen}, {Reach}, {Rebolo}, {Reinecke}, {Remazeilles}, {Renault},
  {Renzi}, {Ristorcelli}, {Rocha}, {Rosset}, {Rossetti}, {Roudier}, {Rozo},
  {Rubi{\~n}o-Mart{\'\i}n}, {Rumsey}, {Rusholme}, {Rykoff}, {Sandri}, {Santos},
  {Saunders}, {Savelainen}, {Savini}, {Schammel}, {Scott}, {Seiffert},
  {Shellard}, {Shimwell}, {Spencer}, {Stanford}, {Stern}, {Stolyarov},
  {Stompor}, {Streblyanska}, {Sudiwala}, {Sunyaev}, {Sutton}, {Suur-Uski},
  {Sygnet}, {Tauber}, {Terenzi}, {Toffolatti}, {Tomasi}, {Tramonte},
  {Tristram}, {Tucci}, {Tuovinen}, {Umana}, {Valenziano}, {Valiviita}, {Van
  Tent}, {Vielva}, {Villa}, {Wade}, {Wandelt}, {Wehus}, {White}, {Wright},
  {Yvon}, {Zacchei}, \& {Zonca}}]{PSZ2}
{Planck Collaboration}, {Ade}, P.~A.~R., {Aghanim}, N., {et~al.} 2016, \aap,
  594, A27

\bibitem[{{Pracy} {et~al.}(2016){Pracy}, {Ching}, {Sadler}, {Croom}, {Baldry},
  {Bland-Hawthorn}, {Brough}, {Brown}, {Couch}, {Davis}, {Drinkwater},
  {Hopkins}, {Jarvis}, {Jelliffe}, {Jurek}, {Loveday}, {Pimbblet}, {Prescott},
  {Wisnioski}, \& {Woods}}]{Pracy2016}
{Pracy}, M.~B., {Ching}, J. H.~Y., {Sadler}, E.~M., {et~al.} 2016, \mnras, 460,
  2

\bibitem[{{Predehl} {et~al.}(2021){Predehl}, {Andritschke}, {Arefiev},
  {Babyshkin}, {Batanov}, {Becker}, {B{\"o}hringer}, {Bogomolov}, {Boller},
  {Borm}, {Bornemann}, {Br{\"a}uninger}, {Br{\"u}ggen}, {Brunner}, {Brusa},
  {Bulbul}, {Buntov}, {Burwitz}, {Burkert}, {Clerc}, {Churazov}, {Coutinho},
  {Dauser}, {Dennerl}, {Doroshenko}, {Eder}, {Emberger}, {Eraerds},
  {Finoguenov}, {Freyberg}, {Friedrich}, {Friedrich}, {F{\"u}rmetz},
  {Georgakakis}, {Gilfanov}, {Granato}, {Grossberger}, {Gueguen}, {Gureev},
  {Haberl}, {H{\"a}lker}, {Hartner}, {Hasinger}, {Huber}, {Ji}, {Kienlin},
  {Kink}, {Korotkov}, {Kreykenbohm}, {Lamer}, {Lomakin}, {Lapshov}, {Liu},
  {Maitra}, {Meidinger}, {Menz}, {Merloni}, {Mernik}, {Mican}, {Mohr},
  {M{\"u}ller}, {Nandra}, {Nazarov}, {Pacaud}, {Pavlinsky}, {Perinati},
  {Pfeffermann}, {Pietschner}, {Ramos-Ceja}, {Rau}, {Reiffers}, {Reiprich},
  {Robrade}, {Salvato}, {Sanders}, {Santangelo}, {Sasaki}, {Scheuerle},
  {Schmid}, {Schmitt}, {Schwope}, {Shirshakov}, {Steinmetz}, {Stewart},
  {Str{\"u}der}, {Sunyaev}, {Tenzer}, {Tiedemann}, {Tr{\"u}mper}, {Voron},
  {Weber}, {Wilms}, \& {Yaroshenko}}]{Predehl2021}
{Predehl}, P., {Andritschke}, R., {Arefiev}, V., {et~al.} 2021, \aap, 647, A1

\bibitem[{{Rengelink} {et~al.}(1997){Rengelink}, {Tang}, {de Bruyn}, {Miley},
  {Bremer}, {Roettgering}, \& {Bremer}}]{WENSS1997}
{Rengelink}, R.~B., {Tang}, Y., {de Bruyn}, A.~G., {et~al.} 1997, \aaps, 124,
  259

\bibitem[{{Robitaille} \& {Bressert}(2012)}]{aplpy2012}
{Robitaille}, T. \& {Bressert}, E. 2012, {APLpy: Astronomical Plotting Library
  in Python}, Astrophysics Source Code Library, record ascl:1208.017

\bibitem[{{Rowan-Robinson} {et~al.}(2013){Rowan-Robinson}, {Gonzalez-Solares},
  {Vaccari}, \& {Marchetti}}]{RR2013}
{Rowan-Robinson}, M., {Gonzalez-Solares}, E., {Vaccari}, M., \& {Marchetti}, L.
  2013, \mnras, 428, 1958

\bibitem[{{Rykoff} {et~al.}(2014){Rykoff}, {Rozo}, {Busha}, {Cunha},
  {Finoguenov}, {Evrard}, {Hao}, {Koester}, {Leauthaud}, {Nord}, {Pierre},
  {Reddick}, {Sadibekova}, {Sheldon}, \& {Wechsler}}]{Rykoff2014}
{Rykoff}, E.~S., {Rozo}, E., {Busha}, M.~T., {et~al.} 2014, \apj, 785, 104

\bibitem[{{Sabater} {et~al.}(2021){Sabater}, {Best}, {Tasse}, {Hardcastle},
  {Shimwell}, {Nisbet}, {Jelic}, {Callingham}, {R{\"o}ttgering}, {Bonato},
  {Bondi}, {Ciardi}, {Cochrane}, {Jarvis}, {Kondapally}, {Koopmans},
  {O'Sullivan}, {Prandoni}, {Schwarz}, {Smith}, {Wang}, {Williams}, \&
  {Zaroubi}}]{Sabater2021}
{Sabater}, J., {Best}, P.~N., {Tasse}, C., {et~al.} 2021, \aap, 648, A2

\bibitem[{{Safouris} {et~al.}(2009){Safouris}, {Subrahmanyan}, {Bicknell}, \&
  {Saripalli}}]{Safouris2009}
{Safouris}, V., {Subrahmanyan}, R., {Bicknell}, G.~V., \& {Saripalli}, L. 2009,
  \mnras, 393, 2

\bibitem[{{Saripalli} {et~al.}(2005){Saripalli}, {Hunstead}, {Subrahmanyan}, \&
  {Boyce}}]{Saripalli2005}
{Saripalli}, L., {Hunstead}, R.~W., {Subrahmanyan}, R., \& {Boyce}, E. 2005,
  \aj, 130, 896

\bibitem[{{Schlafly} {et~al.}(2019){Schlafly}, {Meisner}, \&
  {Green}}]{Schlafly2019}
{Schlafly}, E.~F., {Meisner}, A.~M., \& {Green}, G.~M. 2019, \apjs, 240, 30

\bibitem[{{Schoenmakers} {et~al.}(2000{\natexlab{a}}){Schoenmakers}, {de
  Bruyn}, {R{\"o}ttgering}, {van der Laan}, \& {Kaiser}}]{Schoenmakers2000a}
{Schoenmakers}, A.~P., {de Bruyn}, A.~G., {R{\"o}ttgering}, H.~J.~A., {van der
  Laan}, H., \& {Kaiser}, C.~R. 2000{\natexlab{a}}, \mnras, 315, 371

\bibitem[{{Schoenmakers} {et~al.}(2000{\natexlab{b}}){Schoenmakers}, {Mack},
  {de Bruyn}, {R{\"o}ttgering}, {Klein}, \& {van der Laan}}]{Schoenmakers2000}
{Schoenmakers}, A.~P., {Mack}, K.~H., {de Bruyn}, A.~G., {et~al.}
  2000{\natexlab{b}}, \aaps, 146, 293

\bibitem[{{Sebastian} {et~al.}(2018){Sebastian}, {Ishwara-Chandra}, {Joshi}, \&
  {Wadadekar}}]{Sebastian2018}
{Sebastian}, B., {Ishwara-Chandra}, C.~H., {Joshi}, R., \& {Wadadekar}, Y.
  2018, \mnras, 473, 4926

\bibitem[{{Shabala} {et~al.}(2017){Shabala}, {Deller}, {Kaviraj}, {Middelberg},
  {Turner}, {Ting}, {Allison}, \& {Davis}}]{Shabala2017}
{Shabala}, S.~S., {Deller}, A., {Kaviraj}, S., {et~al.} 2017, \mnras, 464, 4706

\bibitem[{{Shimwell} {et~al.}(2022){Shimwell}, {Hardcastle}, {Tasse}, {Best},
  {R{\"o}ttgering}, {Williams}, {Botteon}, {Drabent}, {Mechev}, {Shulevski},
  {van Weeren}, {Bester}, {Br{\"u}ggen}, {Brunetti}, {Callingham}, {Chy{\.z}y},
  {Conway}, {Dijkema}, {Duncan}, {de Gasperin}, {Hale}, {Haverkorn}, {Hugo},
  {Jackson}, {Mevius}, {Miley}, {Morabito}, {Morganti}, {Offringa}, {Oonk},
  {Rafferty}, {Sabater}, {Smith}, {Schwarz}, {Smirnov}, {O'Sullivan},
  {Vedantham}, {White}, {Albert}, {Alegre}, {Asabere}, {Bacon}, {Bonafede},
  {Bonnassieux}, {Brienza}, {Bilicki}, {Bonato}, {Calistro Rivera}, {Cassano},
  {Cochrane}, {Croston}, {Cuciti}, {Dallacasa}, {Danezi}, {Dettmar}, {Di
  Gennaro}, {Edler}, {En{\ss}lin}, {Emig}, {Franzen}, {Garc{\'\i}a-Vergara},
  {Grange}, {G{\"u}rkan}, {Hajduk}, {Heald}, {Heesen}, {Hoang}, {Hoeft},
  {Horellou}, {Iacobelli}, {Jamrozy}, {Jeli{\'c}}, {Kondapally}, {Kukreti},
  {Kunert-Bajraszewska}, {Magliocchetti}, {Mahatma}, {Ma{\l}ek}, {Mandal},
  {Massaro}, {Meyer-Zhao}, {Mingo}, {Mostert}, {Nair}, {Nakoneczny},
  {Nikiel-Wroczy{\'n}ski}, {Orr{\'u}}, {Pajdosz-{\'S}mierciak}, {Pasini},
  {Prandoni}, {van Piggelen}, {Rajpurohit}, {Retana-Montenegro}, {Riseley},
  {Rowlinson}, {Saxena}, {Schrijvers}, {Sweijen}, {Siewert}, {Timmerman},
  {Vaccari}, {Vink}, {West}, {Wo{\l}owska}, {Zhang}, \& {Zheng}}]{Shimwell2022}
{Shimwell}, T.~W., {Hardcastle}, M.~J., {Tasse}, C., {et~al.} 2022, \aap, 659,
  A1

\bibitem[{{Shimwell} {et~al.}(2019){Shimwell}, {Tasse}, {Hardcastle}, {Mechev},
  {Williams}, {Best}, {R{\"o}ttgering}, {Callingham}, {Dijkema}, {de Gasperin},
  {Hoang}, {Hugo}, {Mirmont}, {Oonk}, {Prandoni}, {Rafferty}, {Sabater},
  {Smirnov}, {van Weeren}, {White}, {Atemkeng}, {Bester}, {Bonnassieux},
  {Br{\"u}ggen}, {Brunetti}, {Chy{\.z}y}, {Cochrane}, {Conway}, {Croston},
  {Danezi}, {Duncan}, {Haverkorn}, {Heald}, {Iacobelli}, {Intema}, {Jackson},
  {Jamrozy}, {Jarvis}, {Lakhoo}, {Mevius}, {Miley}, {Morabito}, {Morganti},
  {Nisbet}, {Orr{\'u}}, {Perkins}, {Pizzo}, {Schrijvers}, {Smith}, {Vermeulen},
  {Wise}, {Alegre}, {Bacon}, {van Bemmel}, {Beswick}, {Bonafede}, {Botteon},
  {Bourke}, {Brienza}, {Calistro Rivera}, {Cassano}, {Clarke}, {Conselice},
  {Dettmar}, {Drabent}, {Dumba}, {Emig}, {En{\ss}lin}, {Ferrari}, {Garrett},
  {G{\'e}nova-Santos}, {Goyal}, {G{\"u}rkan}, {Hale}, {Harwood}, {Heesen},
  {Hoeft}, {Horellou}, {Jackson}, {Kokotanekov}, {Kondapally},
  {Kunert-Bajraszewska}, {Mahatma}, {Mahony}, {Mandal}, {McKean}, {Merloni},
  {Mingo}, {Miskolczi}, {Mooney}, {Nikiel-Wroczy{\'n}ski}, {O'Sullivan},
  {Quinn}, {Reich}, {Roskowi{\'n}ski}, {Rowlinson}, {Savini}, {Saxena},
  {Schwarz}, {Shulevski}, {Sridhar}, {Stacey}, {Urquhart}, {van der Wiel},
  {Varenius}, {Webster}, \& {Wilber}}]{Shimwell2019}
{Shimwell}, T.~W., {Tasse}, C., {Hardcastle}, M.~J., {et~al.} 2019, \aap, 622,
  A1

\bibitem[{{Shulevski} {et~al.}(2019){Shulevski}, {Barthel}, {Morganti},
  {Harwood}, {Brienza}, {Shimwell}, {R{\"o}ttgering}, {White}, {Callingham},
  {Mooney}, \& {Rafferty}}]{Shulevski2019}
{Shulevski}, A., {Barthel}, P.~D., {Morganti}, R., {et~al.} 2019, \aap, 628,
  A69

\bibitem[{{Simonte} {et~al.}(2023){Simonte}, {Andernach}, {Br{\"u}ggen},
  {Best}, \& {Osinga}}]{Simonte2023}
{Simonte}, M., {Andernach}, H., {Br{\"u}ggen}, M., {Best}, P.~N., \& {Osinga},
  E. 2023, \aap, 672, A178

\bibitem[{{Simonte} {et~al.}(2022){Simonte}, {Andernach}, {Br{\"u}ggen},
  {Schwarz}, {Prandoni}, \& {Willis}}]{Simonte2022}
{Simonte}, M., {Andernach}, H., {Br{\"u}ggen}, M., {et~al.} 2022, \mnras, 515,
  2032

\bibitem[{{Srivastava} \& {Singal}(2020)}]{Srivastava2020}
{Srivastava}, S. \& {Singal}, A.~K. 2020, \mnras, 493, 3811

\bibitem[{{Subrahmanyan} {et~al.}(1996){Subrahmanyan}, {Saripalli}, \&
  {Hunstead}}]{Subrahmanyan1996}
{Subrahmanyan}, R., {Saripalli}, L., \& {Hunstead}, R.~W. 1996, \mnras, 279,
  257

\bibitem[{{Subrahmanyan} {et~al.}(2008){Subrahmanyan}, {Saripalli}, {Safouris},
  \& {Hunstead}}]{Subrahmanyan2008}
{Subrahmanyan}, R., {Saripalli}, L., {Safouris}, V., \& {Hunstead}, R.~W. 2008,
  \apj, 677, 63

\bibitem[{{Takey} {et~al.}(2014){Takey}, {Schwope}, \& {Lamer}}]{Takey2014}
{Takey}, A., {Schwope}, A., \& {Lamer}, G. 2014, \aap, 564, A54

\bibitem[{{Tasse} {et~al.}(2021){Tasse}, {Shimwell}, {Hardcastle},
  {O'Sullivan}, {van Weeren}, {Best}, {Bester}, {Hugo}, {Smirnov}, {Sabater},
  {Calistro-Rivera}, {de Gasperin}, {Morabito}, {R{\"o}ttgering}, {Williams},
  {Bonato}, {Bondi}, {Botteon}, {Br{\"u}ggen}, {Brunetti}, {Chy{\.z}y},
  {Garrett}, {G{\"u}rkan}, {Jarvis}, {Kondapally}, {Mandal}, {Prandoni},
  {Repetti}, {Retana-Montenegro}, {Schwarz}, {Shulevski}, \&
  {Wiaux}}]{Tasse2021}
{Tasse}, C., {Shimwell}, T., {Hardcastle}, M.~J., {et~al.} 2021, \aap, 648, A1

\bibitem[{{Tielens} {et~al.}(1979){Tielens}, {Miley}, \&
  {Willis}}]{Tielens1979}
{Tielens}, A.~G.~G.~M., {Miley}, G.~K., \& {Willis}, A.~G. 1979, \aaps, 35, 153

\bibitem[{{Toba} {et~al.}(2019){Toba}, {Yamashita}, {Nagao}, {Wang}, {Ueda},
  {Ichikawa}, {Kawaguchi}, {Akiyama}, {Hsieh}, {Kajisawa}, {Lee}, {Matsuoka},
  {Noboriguchi}, {Onoue}, {Schramm}, {Tanaka}, \& {Komiyama}}]{Toba2019}
{Toba}, Y., {Yamashita}, T., {Nagao}, T., {et~al.} 2019, \apjs, 243, 15

\bibitem[{{Trouille} {et~al.}(2008){Trouille}, {Barger}, {Cowie}, {Yang}, \&
  {Mushotzky}}]{Trouille2008}
{Trouille}, L., {Barger}, A.~J., {Cowie}, L.~L., {Yang}, Y., \& {Mushotzky},
  R.~F. 2008, \apjs, 179, 1

\bibitem[{{Turner} \& {Shabala}(2015)}]{Turner2015}
{Turner}, R.~J. \& {Shabala}, S.~S. 2015, \apj, 806, 59

\bibitem[{{Turner} {et~al.}(2023){Turner}, {Yates-Jones}, {Shabala}, {Quici},
  \& {Stewart}}]{Turner2023}
{Turner}, R.~J., {Yates-Jones}, P.~M., {Shabala}, S.~S., {Quici}, B., \&
  {Stewart}, G. S.~C. 2023, \mnras, 518, 945

\bibitem[{{van Cappellen} {et~al.}(2022){van Cappellen}, {Oosterloo},
  {Verheijen}, {Adams}, {Adebahr}, {Braun}, {Hess}, {Holties}, {van der Hulst},
  {Hut}, {Kooistra}, {van Leeuwen}, {Loose}, {Morganti}, {Moss}, {Orr{\'u}},
  {Ruiter}, {Schoenmakers}, {Vermaas}, {Wijnholds}, {van Amesfoort}, {Arts},
  {Attema}, {Bakker}, {Bassa}, {Bast}, {Benthem}, {Beukema}, {Blaauw}, {de
  Blok}, {Bouwhuis}, {van den Brink}, {Connor}, {Coolen}, {Damstra}, {van
  Diepen}, {de Goei}, {D{\'e}nes}, {Drost}, {Ebbendorf}, {Frank}, {Gardenier},
  {Gerbers}, {Grange}, {Grit}, {Gunst}, {Gupta}, {Ivashina}, {J{\'o}zsa},
  {Janssen}, {Koster}, {Kruithof}, {Kuindersma}, {Kutkin}, {Lucero}, {Maan},
  {Maccagni}, {van der Marel}, {Mika}, {Morawietz}, {Mulder}, {Mulder},
  {Norden}, {Offringa}, {Oostrum}, {Overeem}, {Paragi}, {Pepping}, {Petroff},
  {Pisano}, {Polatidis}, {Prasad}, {de Reijer}, {Romein}, {Schaap},
  {Schoonderbeek}, {Schulz}, {van der Schuur}, {Sclocco}, {Sluman}, {Smits},
  {Stappers}, {Straal}, {Stuurwold}, {Verstappen}, {Vohl}, {Wierenga},
  {Woestenburg}, {Zanting}, \& {Ziemke}}]{vanCappellen2022}
{van Cappellen}, W.~A., {Oosterloo}, T.~A., {Verheijen}, M.~A.~W., {et~al.}
  2022, \aap, 658, A146

\bibitem[{{van der Walt} {et~al.}(2011){van der Walt}, {Colbert}, \&
  {Varoquaux}}]{numpy2011}
{van der Walt}, S., {Colbert}, S.~C., \& {Varoquaux}, G. 2011, Computing in
  Science and Engineering, 13, 22

\bibitem[{{van Haarlem} {et~al.}(2013){van Haarlem}, {Wise}, {Gunst}, {Heald},
  {McKean}, {Hessels}, {de Bruyn}, {Nijboer}, {Swinbank}, {Fallows},
  {Brentjens}, {Nelles}, {Beck}, {Falcke}, {Fender}, {H{\"o}randel},
  {Koopmans}, {Mann}, {Miley}, {R{\"o}ttgering}, {Stappers}, {Wijers},
  {Zaroubi}, {van den Akker}, {Alexov}, {Anderson}, {Anderson}, {van Ardenne},
  {Arts}, {Asgekar}, {Avruch}, {Batejat}, {B{\"a}hren}, {Bell}, {Bell}, {van
  Bemmel}, {Bennema}, {Bentum}, {Bernardi}, {Best}, {B{\^\i}rzan}, {Bonafede},
  {Boonstra}, {Braun}, {Bregman}, {Breitling}, {van de Brink}, {Broderick},
  {Broekema}, {Brouw}, {Br{\"u}ggen}, {Butcher}, {van Cappellen}, {Ciardi},
  {Coenen}, {Conway}, {Coolen}, {Corstanje}, {Damstra}, {Davies}, {Deller},
  {Dettmar}, {van Diepen}, {Dijkstra}, {Donker}, {Doorduin}, {Dromer}, {Drost},
  {van Duin}, {Eisl{\"o}ffel}, {van Enst}, {Ferrari}, {Frieswijk}, {Gankema},
  {Garrett}, {de Gasperin}, {Gerbers}, {de Geus}, {Grie{\ss}meier}, {Grit},
  {Gruppen}, {Hamaker}, {Hassall}, {Hoeft}, {Holties}, {Horneffer}, {van der
  Horst}, {van Houwelingen}, {Huijgen}, {Iacobelli}, {Intema}, {Jackson},
  {Jelic}, {de Jong}, {Juette}, {Kant}, {Karastergiou}, {Koers}, {Kollen},
  {Kondratiev}, {Kooistra}, {Koopman}, {Koster}, {Kuniyoshi}, {Kramer},
  {Kuper}, {Lambropoulos}, {Law}, {van Leeuwen}, {Lemaitre}, {Loose}, {Maat},
  {Macario}, {Markoff}, {Masters}, {McFadden}, {McKay-Bukowski}, {Meijering},
  {Meulman}, {Mevius}, {Middelberg}, {Millenaar}, {Miller-Jones}, {Mohan},
  {Mol}, {Morawietz}, {Morganti}, {Mulcahy}, {Mulder}, {Munk}, {Nieuwenhuis},
  {van Nieuwpoort}, {Noordam}, {Norden}, {Noutsos}, {Offringa}, {Olofsson},
  {Omar}, {Orr{\'u}}, {Overeem}, {Paas}, {Pandey-Pommier}, {Pandey}, {Pizzo},
  {Polatidis}, {Rafferty}, {Rawlings}, {Reich}, {de Reijer}, {Reitsma},
  {Renting}, {Riemers}, {Rol}, {Romein}, {Roosjen}, {Ruiter}, {Scaife}, {van
  der Schaaf}, {Scheers}, {Schellart}, {Schoenmakers}, {Schoonderbeek},
  {Serylak}, {Shulevski}, {Sluman}, {Smirnov}, {Sobey}, {Spreeuw}, {Steinmetz},
  {Sterks}, {Stiepel}, {Stuurwold}, {Tagger}, {Tang}, {Tasse}, {Thomas},
  {Thoudam}, {Toribio}, {van der Tol}, {Usov}, {van Veelen}, {van der Veen},
  {ter Veen}, {Verbiest}, {Vermeulen}, {Vermaas}, {Vocks}, {Vogt}, {de Vos},
  {van der Wal}, {van Weeren}, {Weggemans}, {Weltevrede}, {White}, {Wijnholds},
  {Wilhelmsson}, {Wucknitz}, {Yatawatta}, {Zarka}, {Zensus}, \& {van
  Zwieten}}]{LOFAR2013}
{van Haarlem}, M.~P., {Wise}, M.~W., {Gunst}, A.~W., {et~al.} 2013, \aap, 556,
  A2

\bibitem[{{Von Der Linden} {et~al.}(2007){Von Der Linden}, {Best}, {Kauffmann},
  \& {White}}]{Linden2007}
{Von Der Linden}, A., {Best}, P.~N., {Kauffmann}, G., \& {White}, S. D.~M.
  2007, \mnras, 379, 867

\bibitem[{{Wegner} {et~al.}(2003){Wegner}, {Bernardi}, {Willmer}, {da Costa},
  {Alonso}, {Pellegrini}, {Maia}, {Chaves}, \& {Rit{\'e}}}]{Wegner2003}
{Wegner}, G., {Bernardi}, M., {Willmer}, C.~N.~A., {et~al.} 2003, \aj, 126,
  2268

\bibitem[{{Wen} \& {Han}(2015)}]{Wen2015}
{Wen}, Z.~L. \& {Han}, J.~L. 2015, \apj, 807, 178

\bibitem[{{Wen} \& {Han}(2018)}]{Wen2018b}
{Wen}, Z.~L. \& {Han}, J.~L. 2018, \mnras, 481, 4158

\bibitem[{{Wen} \& {Han}(2021)}]{Wen2021}
{Wen}, Z.~L. \& {Han}, J.~L. 2021, \mnras, 500, 1003

\bibitem[{{Wen} {et~al.}(2018){Wen}, {Han}, \& {Yang}}]{Wen2018a}
{Wen}, Z.~L., {Han}, J.~L., \& {Yang}, F. 2018, \mnras, 475, 343

\bibitem[{{Williams} {et~al.}(2018){Williams}, {Calistro Rivera}, {Best},
  {Hardcastle}, {R{\"o}ttgering}, {Duncan}, {de Gasperin}, {Jarvis}, {Miley},
  {Mahony}, {Morabito}, {Nisbet}, {Prandoni}, {Smith}, {Tasse}, \&
  {White}}]{Williams2018}
{Williams}, W.~L., {Calistro Rivera}, G., {Best}, P.~N., {et~al.} 2018, \mnras,
  475, 3429

\bibitem[{{Williams} {et~al.}(2021){Williams}, {de Gasperin}, {Hardcastle},
  {van Weeren}, {Tasse}, {Shimwell}, {Best}, {Bonato}, {Bondi}, {Br{\"u}ggen},
  {R{\"o}ttgering}, \& {Smith}}]{Williams2021}
{Williams}, W.~L., {de Gasperin}, F., {Hardcastle}, M.~J.~H., {et~al.} 2021,
  \aap, 655, A40

\bibitem[{{Willis} {et~al.}(1974){Willis}, {Strom}, \& {Wilson}}]{Willis1974}
{Willis}, A.~G., {Strom}, R.~G., \& {Wilson}, A.~S. 1974, \nat, 250, 625

\bibitem[{{Wright} {et~al.}(2010){Wright}, {Eisenhardt}, {Mainzer}, {Ressler},
  {Cutri}, {Jarrett}, {Kirkpatrick}, {Padgett}, {McMillan}, {Skrutskie},
  {Stanford}, {Cohen}, {Walker}, {Mather}, {Leisawitz}, {Gautier}, {McLean},
  {Benford}, {Lonsdale}, {Blain}, {Mendez}, {Irace}, {Duval}, {Liu}, {Royer},
  {Heinrichsen}, {Howard}, {Shannon}, {Kendall}, {Walsh}, {Larsen}, {Cardon},
  {Schick}, {Schwalm}, {Abid}, {Fabinsky}, {Naes}, \& {Tsai}}]{Wright2010}
{Wright}, E.~L., {Eisenhardt}, P. R.~M., {Mainzer}, A.~K., {et~al.} 2010, \aj,
  140, 1868

\bibitem[{{Wu} \& {Shen}(2022)}]{Wu2022}
{Wu}, Q. \& {Shen}, Y. 2022, \apjs, 263, 42

\bibitem[{{Xu} {et~al.}(2020){Xu}, {Dai}, {Huang}, {Wang}, {Cheng}, {Shao},
  {Wu}, {Yang}, {Jing}, {Sawicki}, \& {Liu}}]{Xu2020}
{Xu}, H., {Dai}, Y.~S., {Huang}, J.-S., {et~al.} 2020, \apj, 905, 103

\bibitem[{{Yoon} {et~al.}(2008){Yoon}, {Schawinski}, {Sheen}, {Ree}, \&
  {Yi}}]{Yoon2008}
{Yoon}, J.~H., {Schawinski}, K., {Sheen}, Y.-K., {Ree}, C.~H., \& {Yi}, S.~K.
  2008, \apjs, 176, 414

\bibitem[{{York} {et~al.}(2000){York}, {Adelman}, {Anderson}, {Anderson},
  {Annis}, {Bahcall}, {Bakken}, {Barkhouser}, {Bastian}, {Berman}, {Boroski},
  {Bracker}, {Briegel}, {Briggs}, {Brinkmann}, {Brunner}, {Burles}, {Carey},
  {Carr}, {Castander}, {Chen}, {Colestock}, {Connolly}, {Crocker}, {Csabai},
  {Czarapata}, {Davis}, {Doi}, {Dombeck}, {Eisenstein}, {Ellman}, {Elms},
  {Evans}, {Fan}, {Federwitz}, {Fiscelli}, {Friedman}, {Frieman}, {Fukugita},
  {Gillespie}, {Gunn}, {Gurbani}, {de Haas}, {Haldeman}, {Harris}, {Hayes},
  {Heckman}, {Hennessy}, {Hindsley}, {Holm}, {Holmgren}, {Huang}, {Hull},
  {Husby}, {Ichikawa}, {Ichikawa}, {Ivezi{\'c}}, {Kent}, {Kim}, {Kinney},
  {Klaene}, {Kleinman}, {Kleinman}, {Knapp}, {Korienek}, {Kron}, {Kunszt},
  {Lamb}, {Lee}, {Leger}, {Limmongkol}, {Lindenmeyer}, {Long}, {Loomis},
  {Loveday}, {Lucinio}, {Lupton}, {MacKinnon}, {Mannery}, {Mantsch}, {Margon},
  {McGehee}, {McKay}, {Meiksin}, {Merelli}, {Monet}, {Munn}, {Narayanan},
  {Nash}, {Neilsen}, {Neswold}, {Newberg}, {Nichol}, {Nicinski}, {Nonino},
  {Okada}, {Okamura}, {Ostriker}, {Owen}, {Pauls}, {Peoples}, {Peterson},
  {Petravick}, {Pier}, {Pope}, {Pordes}, {Prosapio}, {Rechenmacher}, {Quinn},
  {Richards}, {Richmond}, {Rivetta}, {Rockosi}, {Ruthmansdorfer}, {Sandford},
  {Schlegel}, {Schneider}, {Sekiguchi}, {Sergey}, {Shimasaku}, {Siegmund},
  {Smee}, {Smith}, {Snedden}, {Stone}, {Stoughton}, {Strauss}, {Stubbs},
  {SubbaRao}, {Szalay}, {Szapudi}, {Szokoly}, {Thakar}, {Tremonti}, {Tucker},
  {Uomoto}, {Vanden Berk}, {Vogeley}, {Waddell}, {Wang}, {Watanabe},
  {Weinberg}, {Yanny}, {Yasuda}, \& {SDSS Collaboration}}]{York2000}
{York}, D.~G., {Adelman}, J., {Anderson}, John~E., J., {et~al.} 2000, \aj, 120,
  1579

\bibitem[{{Zhang} {et~al.}(2022){Zhang}, {D'Amico}, {Senatore}, {Zhao}, \&
  {Cai}}]{Zhang2022}
{Zhang}, P., {D'Amico}, G., {Senatore}, L., {Zhao}, C., \& {Cai}, Y. 2022,
  \jcap, 2022, 036

\bibitem[{{Zhou} {et~al.}(2021){Zhou}, {Newman}, {Mao}, {Meisner}, {Moustakas},
  {Myers}, {Prakash}, {Zentner}, {Brooks}, {Duan}, {Landriau}, {Levi}, {Prada},
  \& {Tarle}}]{Zhou2021}
{Zhou}, R., {Newman}, J.~A., {Mao}, Y.-Y., {et~al.} 2021, \mnras, 501, 3309

\bibitem[{{Zou} {et~al.}(2022){Zou}, {Sui}, {Xue}, {Zhou}, {Ma}, {Zhou}, {Nie},
  {Zhang}, {Feng}, {Shen}, \& {Wang}}]{Zou2022}
{Zou}, H., {Sui}, J., {Xue}, S., {et~al.} 2022, Research in Astronomy and
  Astrophysics, 22, 065001

\bibitem[{{Zou} {et~al.}(2017){Zou}, {Zhou}, {Fan}, {Zhang}, {Zhou}, {Nie},
  {Peng}, {McGreer}, {Jiang}, {Dey}, {Fan}, {He}, {Jiang}, {Lang}, {Lesser},
  {Ma}, {Mao}, {Schlegel}, \& {Wang}}]{Zou2017}
{Zou}, H., {Zhou}, X., {Fan}, X., {et~al.} 2017, \pasp, 129, 064101

\bibitem[{{Zovaro} {et~al.}(2022){Zovaro}, {Riseley}, {Taylor}, {Nesvadba},
  {Galvin}, {Malik}, \& {Kewley}}]{Zovaro2022}
{Zovaro}, H. R.~M., {Riseley}, C.~J., {Taylor}, P., {et~al.} 2022, \mnras, 509,
  4997

\end{thebibliography}
\bibliographystyle{aa}

\begin{appendix}

\section{Images of the LDF-GRG sample}

\begin{figure*}[b]
        \centering
        \includegraphics[width= 1\textwidth]{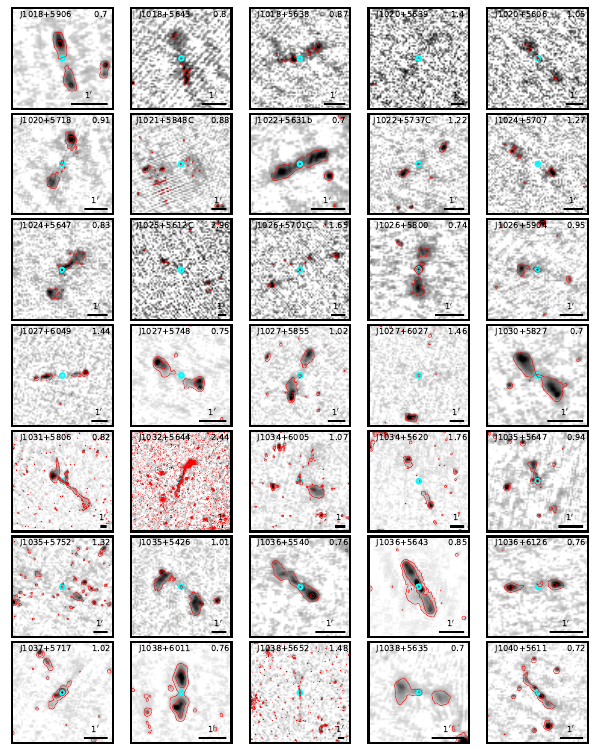}
        \caption{Cutouts of the LOFAR deep fields image around our GRGs at 150 MHz with 3 and 24-$\sigma_{rms}$ red contours superimposed.  The resolution of the images is 6$^{\prime \prime}$. The cyan circle identifies the position of the host galaxy. The bar in the bottom-right corner represents an angular size of 1$^{\prime}$, while the name and the largest linear size (in Mpc) of the GRGs are reported in the top-left and top-right corners, respectively.}
        \label{fig:GRG_images}
  \end{figure*}

\setcounter{figure}{0}

\begin{figure*}
        \centering
        \includegraphics[width= 1\textwidth]{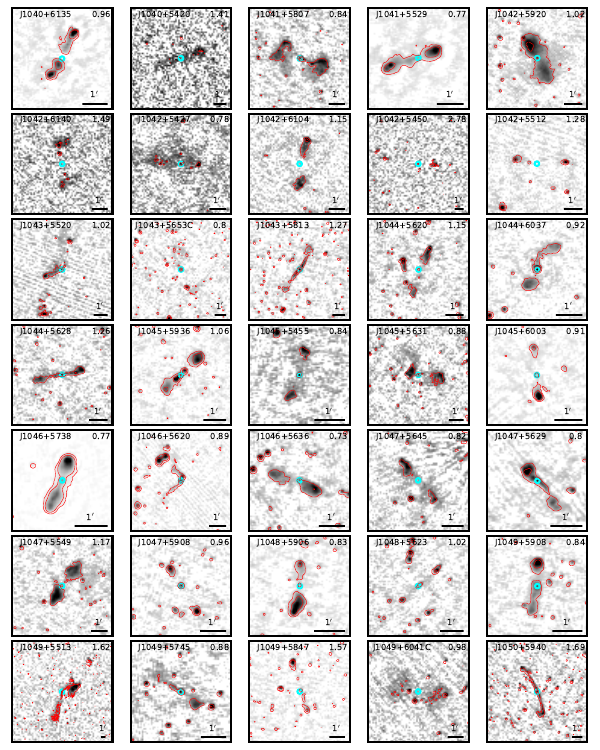}
        \caption{Continued.}
\end{figure*}

\setcounter{figure}{0}

\begin{figure*}
        \centering
        \includegraphics[width= 1\textwidth]{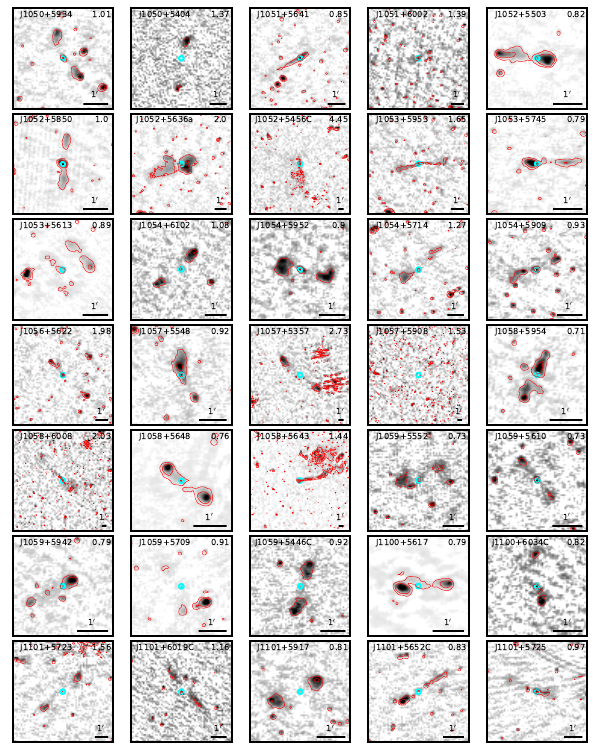}
        \caption{Continued.}
\end{figure*}

\setcounter{figure}{0}

\begin{figure*}
        \centering
        \includegraphics[width= 1\textwidth]{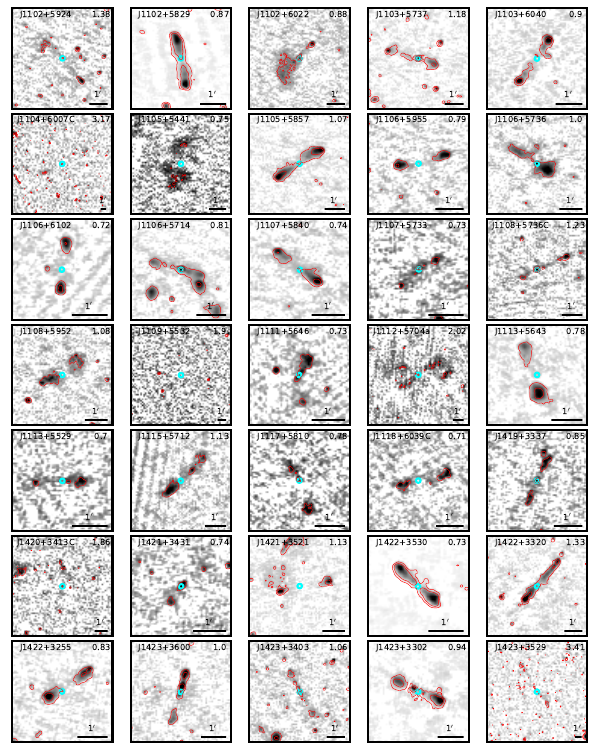}
        \caption{Continued.}
\end{figure*}

\setcounter{figure}{0}

\begin{figure*}
        \centering
        \includegraphics[width= 1\textwidth]{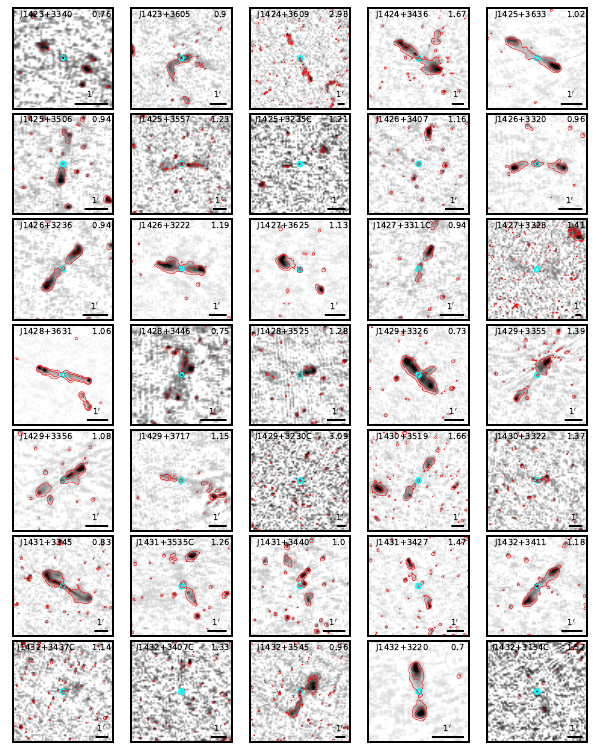}
        \caption{Continued.}
\end{figure*}

\setcounter{figure}{0}

\begin{figure*}
        \centering
        \includegraphics[width= 1\textwidth]{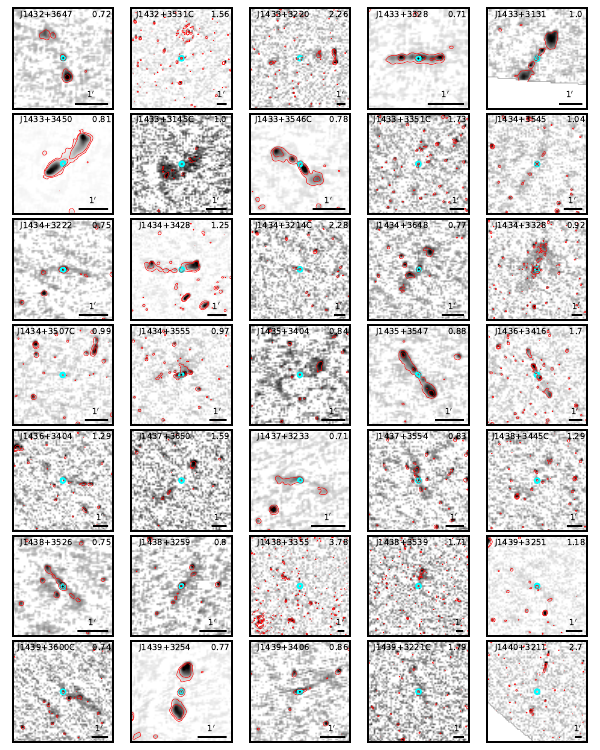}
        \caption{Continued.}
\end{figure*}

\setcounter{figure}{0}

\begin{figure*}
        \centering
        \includegraphics[width= 1\textwidth]{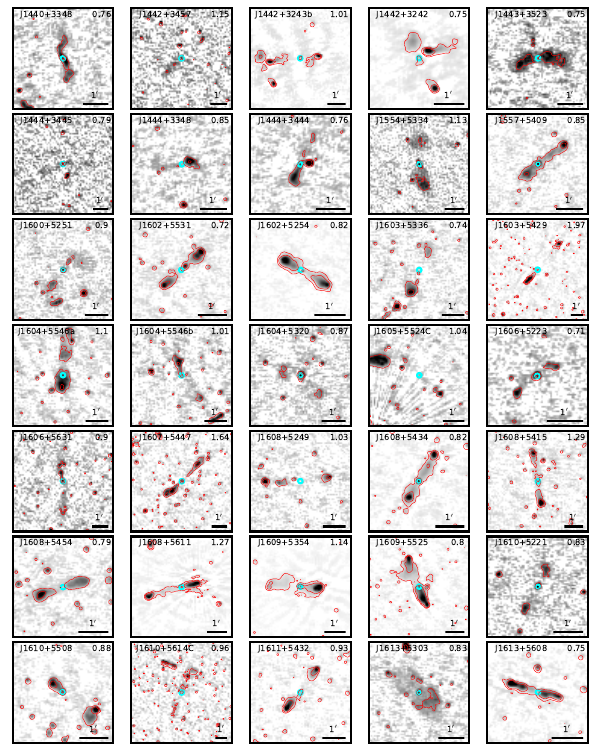}
        \caption{Continued.}
\end{figure*}

\setcounter{figure}{0}

\begin{figure*}
        \centering
        \includegraphics[width= 1\textwidth]{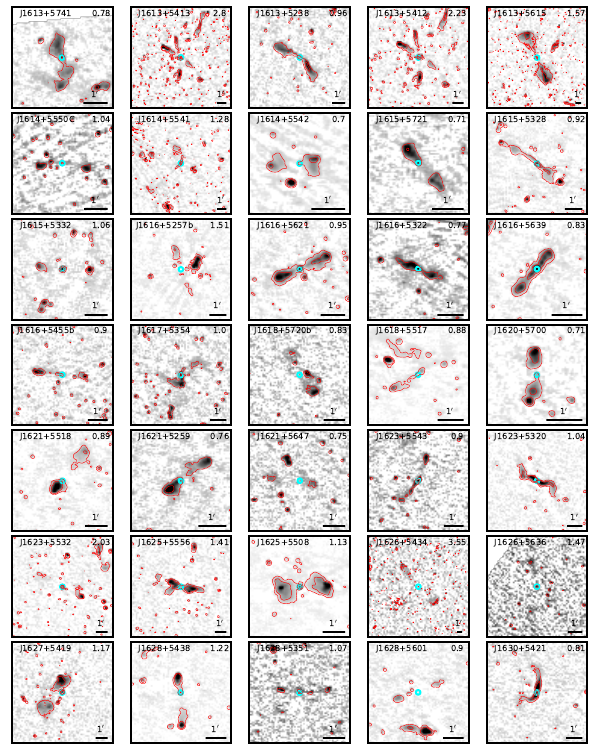}
        \caption{Continued.}
\end{figure*}

\end{appendix}

\end{document}